\documentclass[11pt]{article} 

\usepackage{braun}
\usepackage{multirow}

\begin{document}

\title{Partition Decoupling for Multi-gene Analysis of Gene Expression Profiling Data}

\author{Rosemary Braun$^1$, Gregory Leibon$^2$,  Scott Pauls$^2$, and
Daniel Rockmore$^{2,3}$ \\
{\footnotesize{\textit{$^1$ National Cancer Institute, NIH, Bethesda, MD 20892}}}\\
{\footnotesize{\textit{$^2$ Department of Mathematics, Dartmouth College, Hanover, NH 03755}}}\\
{\footnotesize{\textit{$^3$ The Santa Fe Institute, Santa Fe, NM 87501}}}
}


\date{\today}

\maketitle

\begin{abstract}
We describe an extension and application of a new unsupervised statistical learning technique, known as the Partition Decoupling Method (PDM), to gene expression microarray data.  This method may be used to classify samples based on multi-gene expression patterns and to identify pathways associated with phenotype, without relying upon the differential expression of individual genes.
  
The PDM uses iterated spectral clustering and scrubbing steps, revealing at each iteration progressively finer structure in the geometry of the data.  Because spectral clustering has the ability to discern clusters that are not linearly separable, its performance is superior to distance- and tree-based classifiers.  After projecting the data onto the cluster centroids and computing the residuals (``scrubbing''), one can repeat the spectral clustering, revealing clusters that were not discernible in the first layer.  These iterations, each of which provide a partition of the data that is decoupled from the others, are carried forward until the structure in the residuals is indistinguishable from noise, preventing over-fitting.

This technique is particularly suitable in the context of gene expression data from complex diseases, where phenotypes are not linearly separable and multi-gene effects are likely to play a role.  Because spectral clustering employs a low-dimension embedding of the data, the combined effect of a large number of genes may be simultaneously considered.  Both the dimensionality of the embedding and the number of clusters are determined from the data, yielding an entirely unsupervised classification method.  Here, we describe the PDM in detail and apply it to three publicly available cancer gene expression data sets.  Our results demonstrate that the PDM is able to distinguish cell types and treatments with higher accuracy than is obtained through other approaches.  By applying the PDM on a pathway--by--pathway basis and searching for pathways that permit unsupervised clustering that accurately matches known sample characteristics, we show how the PDM may be used to find sets of mechanistically-related genes that may play a role in disease.

\end{abstract}


\section*{Introduction}

%
Since their first use nearly fifteen years ago~\cite{SCHE95},
microarray gene expression profiling experiments have become a
ubiquitous tool in the study of disease.  The vast number of gene
transcripts assayed by modern microarrays ($10^5$--$10^6$) has
driven forward our understanding of biological processes tremendously,
both by elucidating mechanisms at play in specific phenotypes and
by revealing previously unknown regulatory mechanisms at play in
all cells.  However, the high-dimensional data produced in these
experiments---often comprising many more variables than samples and
subject to noise---presents analytical challenges. 

The analysis of gene expression data can be broadly
grouped into two categories: the identification of differentially
expressed genes (or gene-sets) between two or more known conditions,
and the unsupervised identification (clustering) of samples or genes
that exhibit similar profiles across the data set.  In the former
case, each gene is tested individually for association with the
phenotype of interest, adjusting at the end for the vast number of
genes probed.  Pre-identified gene sets, such as those fulfilling
a common biological function, may then be tested for an overabundance
of differentially expressed genes (e.g., using gene set enrichment
analysis~\cite{GSEA05}); this approach aids biological interpretability
and improves the reproducibility of findings between microarray
studies.
In clustering, the hypothesis that functionally related genes and/or
phenotypically similar samples will display correlated gene expression
patterns motivates the search for groups of genes or samples with 
similar expression patterns. The most
commonly used algorithms are hierarchical clustering~\cite{EISE98},
$k$-means clustering~\cite{HART79,TAVA99} and Self Organizing
Maps~\cite{TAMA99}.  A brief overview may be found in~\cite{DHAE05}.
Of these, $k$-means appears to perform the best~\cite{DHAE05,DATT03,DESO08}.
Relatedly, gene shaving~\cite{HAST00} searches for clusters of genes
showing both high variation across the samples and correlation
across the genes, and several biclustering algorithms (such as~\cite{LI08})
search for class--conditional clusters of correlated genes.
These methods are simple, visually appealing,
and have identified a number of co-regulated genes and phenotype
classes.

While approaches have been fruitful, they also have the potential to miss
causative mechanisms that can be affected by a change in any one of several
genes (such that no single alteration reaches significance) as well
as mechanisms that require the concerted activity of multiple genes
to produce a specific phenotype.  
It is well known that complex diseases, such as cancers, 
exhibit considerable molecular heterogeneity for the
above reasons~\cite{HANA00}.  As a result, individual genes may 
fail to reach significance, and lists of differentially
expressed genes or gene signatures may have poor concordance
across studies.  Additionally, pathway analyses that rely
on single-gene association statistics (such as GSEA~\cite{GSEA05})
may fail to identify causative mechanisms.  For the same reasons, 
clustering algorithms that rely on linearly-separable clusters
(and hence upon differential expression between the clusters)
may fail to partition the samples in a manner that reflects the
true underlying biology.

As an example of how causative genes can be missed in gene--centric
analyses, consider a recent 
study in which gene expression profiles in the Wagyu cattle are
compared to those of the double-muscled Piedmontese cattle~\cite{HUDS09}.
The Piedmontese cattle's muscular hypertrophy is attributable to
a nonfunctional mutation of the myostatin gene (MSTN), but because
MSTN itself is not differentially expressed
between the two bovine models, its biological role cannot be inferred
using traditional analyses of gene expression data.
%
On the other hand, \cite{HUDS09} showed that the functional MSTN
variant was \textit{co}-expressed with its regulatory target MYL2
in Wagyu cattle, whereas the nonfunctional variant in the Piedmontese
cattle did not exhibit co-expression with MYL2.  The correct
identification of this system, in absence of differential expression
at the gene level in MSTN or MYL2, is crucial to understanding the
molecular determinants of the double-muscled phenotype.  This example
serves to underscore the pressing need for analysis methods that
can reveal \textit{systems--level} differences in cases and controls
even when the constituent genes do not exhibit differential expression.


As an alternative approach, we propose here an
analysis technique that is designed to reveal relationships between
samples based on multi-gene expression profiles without requiring
that the genes be differentially expressed (i.e., without requiring
the samples to be linearly separable in the gene-expression space),
and that has the power to reveal relationships between samples at
various scales, permitting the identification of phenotypic 
subtypes.
Our approach adapts a new unsupervised machine-learning technique,
the Partition Decoupling Method (PDM)~\cite{LEIB08,LEIB09}, to gene
expression data.
%
The PDM consists of two iterated components: a spectral clustering
step, in which the correlations between samples in the high-dimensional
feature space is used to partition samples into clusters, followed
by a scrubbing step, in which a projection of the data onto the
cluster centroids is removed so that the residuals may be clustered.
As part of the spectral clustering procedure, a low-dimensional
nonlinear embedding of the data is used; as we will show in the Methods
section,
this both reduces the 
effect of noisy features and permits the partitioning of clusters
with non-convex boundaries.
The clustering and scrubbing steps are iterated until the residuals
are indistinguishable from noise, as determined by comparison to
a resampled null model.  This procedure yields ``layers'' of clusters
that articulate relationships between samples at progressively 
finer scales, and distinguishes the PDM from other clustering
algorithms.


The PDM has a number of satisfying features. The use of spectral
clustering allows identification of clusters that are not necessarily
separable by linear surfaces, permitting the identification of complex
relationships between samples. This means that clusters of samples
can be identified even in situations where the genes do not exhibit
differential expression, 
a trait that makes it particularly well-suited to examining gene
expression profiles of complex diseases.  The PDM employs a low-dimensional
embedding of the feature space, reducing the effect of noise in
microarray studies.  Because the data itself is used to determine
both the optimal number of clusters and the optimal dimensionality
in which the feature space is represented, the PDM provides an entirely
unsupervised method for classification without relying upon heuristics.
Importantly, the use of a resampled null model to determine the 
optimal dimensionality and number of clusters prevents clustering when
the geometric structure of the data is indistinguishable from chance.
By scrubbing the data and repeating the clustering on the residuals, the
PDM permits the resolution of relationships between samples at various scales;
this is a particularly useful feature in the context of gene-expression
analysis, as it permits the discovery of distinct sample subtypes. 
By applying the PDM to gene subsets defined by
common pathways, we can use the PDM to identify gene subsets in which
biologically--meaningful topological structures exist, and infer that
those pathways are related to the clinical characteristics of the
samples (for instance, if the genes in a particular pathway admit
unsupervised PDM partitioning that corresponds to tumor/non-tumor
cell types, one may infer that pathway's involvement in tumorigenesis).
This pathway-based approach has the benefit of incorporating
existing knowledge and being interpretable from a biological
standpoint in a way that searching for sets of highly significant
but mechanistically unrelated genes does not.

A number of other operationally similar, yet functionally distinct,
methods have been considered in the literature.  
%
First, simple spectral clustering has been applied to gene expression data
in~\cite{DESO08}, with mixed success. The PDM improves upon this
both through the use of the resampled null model to provide a
data-driven (rather than heuristic) choice of the clustering
parameters, and by its ability to articulate independent partitions
of the data (in contrast to a single layer) where such structure
is present.  As we will show, these aspects make PDM more powerful than
standard spectral clustering, yielding improved accuracy as well as the
potential to identify sample subtypes that are not already known.
Another novel clustering method is proposed
in \cite{KIM05}, where an adaptive distance norm is used that 
can be shown to identify clusters of different shapes.  The algorithm
iteratively assigns clusters and refines the distance metric scaling
parameter in a cluster-conditional fashion based on each cluster's
geometry. This approach is able
to identify clusters of mixed sizes and shapes that cannot
be discriminated using fixed Euclidean or Mahalanobis distance
metrics, and thus is a considerable improvement over $k$-means
clustering.
However, the method as described in~\cite{KIM05} is computationally
expensive and cannot identify
non-convex clusters as spectral clustering, and hence the PDM, can.  
Alternatively, SPACC~\cite{QUI09} uses the same type of
non-linear embedding of the data as is used in the PDM, which permits
the articulation of non-convex boundaries.  In SPACC~\cite{QUI09},
a single dimension of this embedding is used to recursively partition the data into two
clusters.  The partitioning is carried out until each cluster is
solely comprised of one class of samples,  yielding a classification
tree.  In this way, SPACC may also in some cases permit partitioning 
of known sample
classes into subcategories.  However, SPACC differs from the
PDM in two crucial ways.  First, the PDM's use of a data-determined
number of informative dimensions permits more accurate clusterings
than those obtained from a single dimension in SPACC.  Second, SPACC
is a semi-supervised algorithm that uses the known class labels
to set a stopping threshold.
Because there is no comparison to a null model, as in the
PDM, SPACC will partition the data until the clusters are
pure with respect to the class labels.  This means that
groups of samples with distinct molecular subtypes but identical
class labels will remain unpartitioned (SPACC cannot be used
to reveal novel subclasses) and that groups of samples with differing
class labels but indistinguishable molecular characteristics will
be artificially divided until the purity threshold is reached.  
In this respect, the PDM improves on SPACC by ``letting the data speak.''
A fourth approach, 
QUBIC~\cite{LI08} is a graph theoretic algorithm that
identifies sets of genes with similar class-conditional coexpression
patterns (biclusters) by employing a network representation of the
gene expression data and agglomeratively finding heavy subgraphs
of co-expressed genes. In contrast to the unsupervised clustering
of the PDM, QUBIC is a supervised method that is designed to find
gene subsets with coexpression patterns that differ between pre-defined
sample classes.  In \cite{LI08} it is shown that QUBIC is able to
identify functionally related gene subsets with greater accuracy
than competing biclustering methods; still, QUBIC is only able to
identify biclusters in which the genes show strict correlation or
anticorrelation coexpression patterns, which means that gene sets
with more complex coexpression dynamics cannot be identified.  
The PDM is thus unique in a number of ways: not only is it able to
partition clusters with nonlinear and nonconvex boundaries, it does
so in an unsupervised manner (permitting the identification of
unknown subtypes) and in the context of comparison to a null
distribution that both prevents clustering by chance and reduces
the influence of noisy features.  Moreover, the PDM's iterated
clustering and scrubbing steps permit the identification of independent
(i.e., decoupled) partitions in the data.


In this manuscript, we describe the PDM algorithm and demonstrate its
application to several publicly-available gene-expression data sets.
To illustrate the PDM's ability to articulate independent partitions
of samples, we apply it to genome-wide expression data from a four
phenotype, three exposure radiation response study~\cite{REIG04}.
The PDM partitions the samples by exposure and then by phenotype,
yielding higher accuracy for predictions of radiation sensitivity
than previously reported~\cite{REIG04}.
We also compare the PDM results to those obtained in a recent~\cite{DESO08}
comparison of clustering techniques, demonstrating the PDM's 
ability to identify cancer subtypes from global 
patterns in the gene expression data.
Next, we apply the PDM using gene subsets defined by pathways rather
than the global gene expression data, demonstrating how the PDM
can be used to find biological mechanisms that relate to the 
phenotype of interest.
We demonstrate pathway-PDM in both the radiation response data~\cite{REIG04}
as well as a larger prostate cancer data set~\cite{SING02}.
Our results suggest that
the PDM is a powerful tool for articulating relationships between
samples and for identifying  pathways containing multi-gene expression 
patterns that distinguish phenotypes.

\section*{Methods}

\subsection*{Partition Decoupling Algorithm}

The partition decoupling method (PDM) was first described in~\cite{LEIB08}.  
We summarize it here, and discuss its application to gene-expression data.
The PDM consists of two iterated submethods: the first, spectral
clustering, finds the dominant structures within the system, while
the second ``scrubbing'' step removes this structure such that the
next clustering iteration can distinguish finer-scale relationships
within the residual data.  The two steps are repeated until the
residuals are indistinguishable from noise.  By performing successive
clustering steps, factors contributing to the partitioning of the
data at different scales may be revealed.

\subsubsection*{Spectral Clustering}

The first step, spectral clustering, serves to identify clusters
of samples in high-dimensional gene-expression space.
The motivation is simple: given a set of samples and a measure of
pairwise similarity $s_{i,j}$ between each pair, we wish to partition
the data such that samples within one cluster are similar to each
other based on their gene expression profiles. A summary of the
spectral clustering algorithm is given in Table~\ref{specclus}.


Spectral clustering offers several advantages over traditional
clustering algorithms such as those reviewed in~\cite{DHAE05}. 
Most importantly, no constraint is placed on the geometry of the
data, in contrast to the tree-like structure imposed by hierarchical
clustering~\cite{EISE98} or the requirement that clusters be convex
in the feature space when using distance-based $k$-means
clustering~\cite{HART79,TAVA99} and Self Organizing Maps~\cite{TAMA99}.
Spectral clustering also uses a low-dimensional embedding of the
data, thus excluding the noisy, high-frequency components.

In spectral clustering, the data are represented as a complete graph
in which nodes correspond to samples and edge weights $s_{i,j}$
correspond to some measure of similarity between a pair of nodes $i$
and $j$.  Spectral graph theory (see, e.g., \cite{CHUN97}) is brought to bear to find groups
of connected, high-weight edges that define clusters of samples.  This
problem may be reformulated as a form of the min-cut problem: cutting the graph
across edges with low weights, so as to generate several subgraphs for
which the similarity between nodes is high and the cluster sizes preserve some form of balance in the network.  It has been
demonstrated~\cite{CHUN97,NG2002,LUXB07} that 
solutions to relaxations of these kinds of combinatorial problems (i.e., converting the problem of finding a minimal configuration over a very large collection of discrete samples to achieving an approximation via the solution to a related continuous problem) can be framed as  an
eigendecomposition of a graph Laplacian matrix $\mat{L}$,
In particular, we use the Laplacian matrix formed from the adjacency matrix
$\mat{S}$ (comprised of $s_{i,j}$) and the diagonal degree matrix
$\mat{D}$ with elements $d_i = \sum_j s_{i,j}$:
\begin{equation}
\mat{L} = \mat{I} - \mat{D}^{-1/2} \, \mat{S} \, \mat{D}^{-1/2} \, .
\end{equation}
The similarity measure between two data points is computed (as
in~\cite{LEIB09}) from their correlation $\rho_{i,j}$ by first
converting the correlation to a chord distance on the unit sphere and
then exponentiating,
\begin{equation}
s_{i,j} = \exp \left( \frac{ - \Big( \sin \big( \arccos ( \rho_{i,j})/2 \big) \Big)^2 }{ \sigma ^2} \right) \, ,
\label{sim1}
\end{equation}
where $\sigma$ determines how quickly $s_{i,j}$ falls off with the
correlation $\rho_{i,j}$ and may be tuned to reveal structure at
various scales of the system.

The spectrum of $\mat{L}$ contains information regarding the graph
connectivity.  Specifically, the number of zero-value eigenvalues
corresponds to the number of connected components; since we have a
complete graph, there will be exactly one.  The second-smallest
eigenvalue and its associated eigenvector (the so-called Fiedler
value $\lambda_1$ and vector $v_1$) encodes a coarse geometry of
the data, effectively the coordinates for the ``best'' (in the sense
of clustering) one-dimensional embedding of the network. Successive
eigenvectors enable the articulation of finer resolution.
By embedding the data into a smaller-dimensional space defined by the
low-frequency eigenvectors and clustering the embedded data, the
geometry of the data may be revealed.

The embedded data are then be clustered using
$k$-means~\cite{HART79}.  Because $k$-means clustering is
by nature stochastic~\cite{HART79}, multiple $k$-means runs are
performed and the clustering yielding the smallest within-cluster
sum of squares is chosen.  In order to use $k$-means on the embedded
data, two parameters need to be chosen: the number of eigenvectors
$l$ to use (that is, the dimensionality of the embedded data) and
the number of clusters $k$ into which the data will be clustered.

\paragraph{Optimization of $l$}  
The optimal dimensionality of the embedded data is obtained by
comparing the eigenvalues of the Laplacian to the distribution of
Fiedler values expected from null data.  The motivation of this
approach follows from the observation that the size of eigenvalues
corresponds to the degree of structure (see~\cite{LUXB07}), with
smaller eigenvalues corresponding to greater structure.  Specifically,
we wish to construct a distribution of null Fiedler values---eigenvalues
encoding the coarsest geometry of randomly organized data---and
select the eigenvalues from the true data that are significantly
small with respect to this distribution (below the 0.05 quantile).
In doing so, we select the eigenvalues that indicate greater structure
than would be expected by chance alone.  The idea is that the
distribution of random Fiedler values give a sense of how much
structure we could expect of a comparable random network. We thus
take a collection of perpendicular axes, onto each of which the
projection of the data would reveal more structure than we would
expect at random.

The null distribution of Fiedler values is obtained through resampling
$s_{i,j}$ (preserving $s_{i,j}=s_{j,i}$ and
$s_{i,i}=1$).  This process may be thought of as ``rewiring'' the
network while retaining the same distribution of edge weights. 
This has the effect of destroying structure by dispersing clusters
(subgraphs containing high edge weights) and creating new clusters
by random chance.  Because the raw data itself is not resampled, the
resulting resampled network is one which has the same marginal gene
expression distributions and gene-gene correlations as the original
data, and is thus a biologically comparable network to that in the
true data.

\comment{Here, the null distribution of Fiedler values is obtained through
row-wise resampling of the data, followed by recomputation of $\mat{L}$.  By
resampling the data within rows, correlation between the samples
(columns) is destroyed while preserving the marginal distributions
of the gene expressions (rows).  By performing this resampling
multiple times and recomputing the null Fiedler value $\lambda'_1$,
a distribution is obtained; eigenvalues $\lambda$ falling below the
0.05 quantile of the distribution of $\lambda'_1$ are denoted as
significant.  In the case where no non-zero $\lambda$ meets this
threshold, we conclude that there is no structure present in our
data which is distinguishable from noise, and halt the procedure.

Although row-wise resampling preserves the distribution within each
feature, it has two potential drawbacks: first, row-wise resampling
is computationally expensive, and second, it destroys correlations
not only between samples (columns) but also between genes (rows).
An alternative approach involves resampling the similarities $s_{i,j}$
themselves, which may be thought of as ``rewiring'' the network.
Because this approach retains greater similarity to the original
data, the significance threshold obtained is smaller; however, in the
case of the data described here, the clustering results were
identical, suggesting that the relevant structure is contained
within the smaller set of eigenvectors.  While a full comparison
of these approaches is not presented in this paper, it should be
noted that the method for constructing the null distribution of
Fiedler values must be carefully chosen with respect to the underlying
process generating the data.}

\paragraph{Optimization of $k$}
Methods for obtaining the number of clusters $k$ suitable for
partitioning a data set are an open research question (see, e.g.,
\cite{LUXB07,STIL04} and references therein).  Our approach exploits
the property~\cite{LEIB09, LUXB07} that clustering the entries in
the Fiedler vector yields the best decomposition of the network
components.  Consequently, one can use the number peaks in the
density of the Fiedler vector---that is, the number of values about
which the elements of $v_1$ are clustered---as the number of clusters.
(This procedure is roughly analogous to finding regions of high
density along the first principle component of the data.)  To obtain
this value, we fit a Gaussian mixture model~\cite{MCLA04} with 2--30
components (assuming unequal variances), compute the Bayesian
Information Criterion (BIC) for each mixture model, and choose the
optimum number of components (for details of the implementation,
see \cite{FRAL99,FRAL06}).

\comment{
Methods for obtaining the number of clusters $k$ suitable for
partitioning a data set are an open research question (see, e.g.,
\cite{LUXB07,STIL04} and references therein).  We describe
our approach here, along with some alternatives, noting that there is no
one-size-fits-all solution.

The first approach exploits the property~\cite{LUXB07, LEIB09} that
clustering the entries in the Fiedler vector yields the best
decomposition of the network components.  Consequently, one can use
the number peaks in the density of the Fiedler vector---that is, the
number of values about which the elements of $v_1$ are clustered---as
the number of clusters.  (This is procedure is roughly
analogous to finding regions of high density along the first principle
component of the data.)  To obtain this value, we fit a Gaussian
mixture model~\cite{MCLA04} with 2--30 components (assuming unequal
variances), compute the Bayesian Information Criterion (BIC) for
each mixture model, and choose the optimum number of components
(for details of the implementation, see \cite{FRAL99,FRAL06}).

An alternative approach attempts to optimize the robustness of the
clustering with respect to increasing $k$.  The underlying idea
here is that, for the correct number of clusters, incrementing to
$k+1$ clusters will leave the majority of the samples in the $k$
clusters as previously obtained, assigning only outliers to the new
cluster.  In practice, this is done by computing the clustering for
$k=$2--30 clusters, keeping track for each $k$ of how many additional
clusters could be added before one of the original clusters split
such that no new cluster has $>$50\% of the data of the original
cluster.  The $k$ for which this statistic is at a maximum is then
considered optimal.

A related approach takes advantage of the expectation that, at the
correct value of $k$, the clustering will be stable even with a
subset of the data, whereas if $k$ is too large, the clustering
will fluctuate strongly with small differences in the data.  In this 
approach, one splits the rows of the data matrix at random into two subsets
and clusters the columns for each; the clusters are then compared between
the two subsets.  By performing this split-and-compare procedure multiple
times for a range of $k$, the $k$ may be optimized to reduce the 
discrepancy between the clusterings of the split data.

Each of these approaches was applied, with similar results, to the
data presented here; we present in detail the results using the BIC-based
method.
}

\vspace{1em}\noindent Once $k$ and $l$ have been assigned, the data embedded
in the $l$ eigenvectors is clustered using $k$-means~\cite{HART79}.
The spectral clustering procedure offers several advantages over
simple clustering of the original data using $k$-means: first, the
Fiedler vector provides a natural means to estimate the number of
clusters; and second, because spectral clustering operates on
similarity of the samples, rather than planar cuts of the
high-dimensional feature space, complex correlation structures can
be identified.  A complete discussion of the advantages of spectral
clustering is given in~\cite{CHUN97,NG2002,LUXB07}.

To illustrate the power of this method, let consider a toy data set
called ``two\_circles'' in which 200 data points are placed in two
dimensional space in two concentric circles, as depicted in
Fig.~\ref{twocircs}. 
\comment{Patterns of this type will arise when two oscillatory genes,
such as those involved in circadian rhythms, are sampled from a
population that has not been synchronized (as is the case with most
human data sets).  The radii of the co-expression circles
will be dictated by the amplitude of the gene oscillations. 
While the data here is simulated, we note the
biological relevance of this example: in~\cite{PTIT07}
it is found that the majority of mammalian genes oscillate, and---as
observed in~\cite{PTIT07} and elsewhere---the amplitude of oscillatory
genes differs between tissue types and is associated with the gene's
function.  These observations led to the conclusion in~\cite{PTIT07}
that pathways should be considered as dynamic systems of genes 
oscillating in coordination with each other, and underscores the need
to detect amplitude differences in co-oscillatory genes as depicted in
Fig.~\ref{twocircs}.}
Because $k$-means alone can only identify
clusters with convex hulls, $k$-means clustering using $k=2$ produces
an arbitrary, linear division of the data as shown in Fig.~\ref{twocircs}(a).
In contrast, spectral clustering identifies the two rings
as individual clusters, as seen in Fig.~\ref{twocircs}(b).
While $k$-means took $k=2$ as an input from the user, the spectral
clustering example determined $k=2$ from the data, as shown in
Fig.~\ref{twocircs}(c); the rug plot depicts the distribution
of the Fiedler vector coordinates, in which two peaks are readily visible and
chosen as indicative of two clusters, as described above.

While the two\_circles data is simulated, we note that patterns of
this type will arise when out-of-phase oscillatory genes, such as those
involved in circadian rhythms or cell cycle processes, are
sampled; the radii of the co-expression circles
will be dictated by the amplitude of the gene oscillations.
An illustration of such patterns in nature is provided in 
Fig.~\ref{spellcyc}, which depicts the co-expression pattern of three
cell-cycle related genes in CDC-28 and elutriation synchronized yeast
cells from~\cite{SPEL98}.  The elutriation synchronized cells exhibit
much smaller amplitude oscillations than do the CDC-28 synchronized cells; 
while the CDC-28 and elutriation synchronized cells cannot
be distinguished using k-means, the distinction is readily made via
spectral clustering.  The biological relevance of patterns such as 
those depicted in Figs.~\ref{twocircs} and~\ref{spellcyc} has been
noted in mammalian systems as well; in~\cite{PTIT07}
it is found that the majority of mammalian genes oscillate and
that the amplitude of oscillatory
genes differs between tissue types and is associated with the gene's
function.  These observations led to the conclusion in~\cite{PTIT07}
that pathways should be considered as dynamic systems of genes
oscillating in coordination with each other, and underscores the need
to detect amplitude differences in co-oscillatory genes as depicted in
Figs.~\ref{twocircs} and~\ref{spellcyc}.

The benefit of spectral clustering for pathway-based analysis in
comparison to over-representation analyses such as GSEA~\cite{GSEA05}
is also evident from the two\_circles example in Fig.~\ref{twocircs}.
Let us consider a situation in which the $x$-axis represents the
expression level of one gene, and the $y$-axis represents another;
let us further assume that the inner ring is known to correspond
to samples of one phenotype, and the outer ring to another.  A
situation of this type may arise from differential misregulation
of the $x$ and $y$ axis genes.  However, while the variance in the
$x$-axis gene differs between the ``inner'' and ``outer'' phenotype,
the means are the same ($0$ in this example); likewise for the
$y$-axis gene.  In the typical single-gene $t$-test analysis of
this example data, we would conclude that neither the $x$-axis nor
the $y$-axis gene was differentially expressed; if our gene set
consisted of the $x$-axis and $y$-axis gene together, it would not
appear as significant in GSEA~\cite{GSEA05}, which measures an
abundance of single-gene associations.  Yet, unsupervised spectral
clustering of the data would produce categories that correlate
exactly with the phenotype, and from this we would conclude that a
gene set consisting of the $x$-axis and $y$-axis genes plays a role
in the phenotypes of interest.  We exploit this property in applying
the PDM by pathway to discover gene sets that permit the accurate
classification of samples.

\subsubsection*{Scrubbing}

After the clustering step has been performed and each data point
assigned to a cluster, we wish to ``scrub out'' the portion of the
data explained by those clusters and consider the remaining variation.
This is done by computing first the cluster centroids (that is, the
mean of all the datapoints assigned to a given cluster), and then
subtracting the data's projection onto each of the centroids from
the data itself, yielding the residuals.  The clustering step may
then be repeated on the residual data, revealing structure that may
exist at multiple levels, until either a) the eigenvalues of the
Laplacian in the scrubbed data are indistinguishable from those of
the resampled graph as described above; or b) the cluster centroids
are linearly dependent.  (It should be noted here that the residuals
may still be computed in the latter case, but it is unclear how to
interpret linearly dependent centroids.)

\subsection*{Implementation}
The PDM as described above was implemented in R~\cite{R} and applied
to the following data sets.  Genes with missing expression values
were excluded when computing the (Pearson) correlation $\rho_{i,j}$
between samples. In the $l$-optimization step, 60 resamplings of
the correlation coefficients were used to determine the dimension
of the embedding $l$.  In the clustering step, 30 $k$-means runs
were performed, choosing the clustering yielding the smallest
within-cluster sum of squares.

\subsection*{Data}
\paragraph{Radiation Response Data}
These data come from a gene-expression profiling study of radiation
toxicity designed to identify the determinants of adverse reaction
to radiation therapy~\cite{REIG04}.  In this study, skin fibroblasts
from 14 patients with high radiation sensitivity (High-RS) were collected
and cultured, along with those from three control groups: 13 patients 
with low radiation-sensitivity (Low-RS), 15 healthy individuals, and
15 individuals with skin cancer.  The cells were then subject to
mock (M), ultraviolet (U) and ionizing (I) radiation exposures.
As reported in~\cite{REIG04}, RNA from these 171 samples comprising
four phenotypes and three treatments were hybridized to Affymetrix
HGU95AV2 chips, providing gene expression data for each sample for
12615 unique probes.  The microarray data was normalized using
RMA~\cite{RMA}.  The gene expression data is publicly available and
was retrieved from the Gene Expression Omnibus~\cite{GEO} repository
under record number GDS968.

\paragraph{DeSouto Multi-study Benchmark Data}
These data comprise filtered gene expression levels from 21 cancer
studies using Affymetrix microarrays along with associated class
labels.  The data were analyzed previously in~\cite{DESO08}, where
several clustering methods were applied to compare algorithmic
performance.  The data were obtained from their original sources
and subjected to filtering as described in~\cite{DESO08}; we obtained
the filtered sets as used in~\cite{DESO08} and made available by
the authors.  This permits a direct comparison of the PDM results
to those reported in~\cite{DESO08}.

\paragraph{Singh Prostate Data}
These data come from a gene-expression profiling study of prostate
cancer comprising 52 tumor samples (T) and 50 tumor-adjacent normal
samples (N) from 52 men who had undergone radical
prostatectomy~\cite{SING02}.  RNA was hybridized to Affymetrix
HGU95AV2 chips, providing gene expression data for each sample for
12615 unique probes.  The microarray data CEL files were downloaded
from the Broad Institute website and normalized using RMA~\cite{RMA}.

\comment{
\paragraph{Yu Prostate Data}
These data come from a gene-expression profiling study of normal,
stromal, tumor, and metastatic prostate tissue~\cite{CHAN07,YU04}.
The gene expression data is publicly available through the Gene
Expression Omnibus~\cite{GEO} repository under record number GDS2545.
This data consisted of 18 normal prostate samples from organ donors,
65 prostate tumor samples, 25 prostate cancer metastasis samples,
and 63 tumor-adjacent normal (stromal) prostate samples hybridized
to Affymetrix HGU95A chips.  While these data were normalized with
respect to one-another using MAS5 [Affymetrix Corporation, Santa
Clara, Ca], they were not normalized with respect to the Singh
prostate data.
}

\paragraph{Pathway annotation}
The BioConductor~\cite{GENT04} annotation packages hgu95av2.db, hgu95a.db,
and KEGG.db were used to map Affymetrix probe IDs to KEGG pathways.  Only
KEGG pathways were investigated.  A total of 203 KEGG pathways containing
genes probed in the above data were identified.

\section*{Results}

We apply the PDM to several cancer gene expression data sets to
demonstrate how it may be used to reveal multiple layers of structure.
In the first data set~\cite{REIG04}, the PDM articulates two independent partitions
corresponding to cell type and cell exposure, respectively.  The
second data set~\cite{DESO08} demonstrates how successive partitioning by the PDM
can reveal disease and tissue subtypes in an unsupervised way.  We
then carry out Pathway-PDM to identify the biological mechanisms
that drive phenotype-associated partitions.  In addition to applying
it to the radiation response data set mentioned above~\cite{REIG04}, we also apply
Pathway-PDM to a prostate cancer data set~\cite{SING02}, and briefly discuss
how the Pathway-PDM results show improved concordance of significant
pathways identified in the Singh data~\cite{SING02} with those previously
identified in several other prostate cancer data sets~\cite{MANO06}.

\subsection*{Partition Decoupling in Cancer Gene Expression Data}

\paragraph{Radiation Response Data}
We begin by applying the PDM to the radiation response data~\cite{REIG04}
to illustrate how it may be used to reveal multiple layers of
structure that, in this case, correspond to radiation exposure and 
sensitivity.  In the first layer, spectral clustering classifies the
samples into three groups that correspond precisely to the treatment
type.  The number of clusters was 
obtained using the BIC optimization method as described, and
resampling the correlation coefficients  was used to determine the
dimension of the embedding $l$ using 60 permutations; 30 $k$-means
runs were performed, choosing the clustering yielding the smallest
within-cluster sum of squares.  Classification results are given
in Table~\ref{tab1} and Figure~\ref{pdmfig}(a); the unsupervised 
algorithm correctly identifies that three clusters are present
in the data, and assigns samples to clusters in a manner consistent
with their exposure.

In order to compare the performance of spectral clustering to that
of $k$-means, we ran $k$-means on the original data using $k=3$ and
$k=4$, corresponding to the number of treatment groups and number
of cell type groups respectively.  As with the spectral clustering,
30 random $k$ means starts were used, and the smallest within-cluster
sum of squares was chosen.  The results, given in Tables~\ref{tab2}
and \ref{tab3}, show substantially noisier classification than the
results obtained via spectral clustering.  It should also be noted
that the number of clusters $k$ used here was not derived from the
characteristics of the data, but rather assigned in a supervised
way based on additional knowledge of the probable number of
categories (here, dictated by the study design).
While the pure $k$-means results are noisy, the $k=4$ classification
yields a cluster that is dominated by the highly radiation-sensitive
cells (cluster 4, Table~\ref{tab3}).  Membership in this cluster
versus all others identifies highly radiation-sensitive cells with
62\% sensitivity and 96\% specificity; if we restrict the analysis
to the clinically-relevant comparison between the last two cell
types---that is, cells from cancer patients who show little to no
radiation sensitivity and those from cancer patients who show high
radiation sensitivity---the classification identifies radiation-sensitive
cells with 62\% sensitivity and 82\% specificity.  

The result from the $k=4$ $k$-means classification suggest 
that there exist cell-type specific differences
in gene expression between the high radiation sensitivity cells and
the others.  To investigate this, we perform the ``scrubbing'' step
of the PDM, taking only the residuals of the data after projecting onto
the clusters obtained in the first pass.  
As in the first layer, we use the BIC
optimization method to determine the number of clusters $k$ and
resampling of the correlations to determine the dimension of the
embedding $l$ using 60 permutations.
The second layer of structure revealed by the PDM paritioned the
high-sensitivity samples from the others into two clusters.
Classification results are given in Table~\ref{tab4} and
Figure~\ref{pdmfig}(b), and it can be seen that the partitioning
of the radiation-sensitive samples is highly accurate (83\%
sensitivity; 91\% specificity across all samples, 72\% when comparing
solely to low radiation-sensitivity patient samples).

Further PDM iterations resulted in residuals that were indistinguishable
from noise (see Methods); we thus conclude that there are only two layers of
structure present in the data: the first corresponding to exposure,
and the second to radiation sensitivity.  That is, there exist patterns
in the gene expression space that distinguish UV- and ionizing radiation
exposed cells from mock-treated cells (and from each other), and
that there exist further patterns that distinguish 
high-sensitivity cells from the rest.  Together, these independent
(decoupled) sets of clusters describe six categories, as shown in
Fig.~\ref{pdmfig}(c), wherein the the second layer partitions the
radiation sensitive cells from the others in each exposure-related
partition.  The fact that the mock-exposure as well as the UV- and
IR-exposure partitions are further divided by radiation sensitivity
in the second layer suggests that there exist constitutive differences
in the radiation sensitive cells that distinguish them from the
other groups even in the absence of exposure.  Importantly, the
data-driven methodology of PDM identifies only phenotypic clusters,
corresponding to the high-sensitivity cells and the three control
groups combined, without further subpartitioning the combined
controls.  This suggests that the three control groups do
not exhibit significant differences in their global gene-expression
profiles.

%
In the original analysis of this data~\cite{REIG04}, the authors
used a linear, supervised algorithm (SAM, a nearest shrunken centroids
classifier~\cite{TUSH01}) to develop a predictor for the high-sensitivity
samples.  This approach obtained 64.2\% sensitivity and 100\%
specificity~\cite{REIG04}, yielding a clinically useful predictor.
The PDM's unsupervised detection of the high sensitivity sample
cluster suggests that the accuracy in \cite{REIG04} was not a result
of overfitting to training data; moreover, the PDM's ability to
identify those samples with higher sensitivity (83\%) than in
\cite{REIG04} indicates that there exist patterns of gene expression
distinct to the radiation-sensitive patients which were not identified
in the SAM analysis, but are detectable using the PDM.

\comment{Also as in the pure $k$-means results, no distinction is seen between
the healthy skin fibroblasts and those of skin cancer patients, who
were expected to show altered UV response; patients who had little
to no radiation sensitivity like between the (insensitive) healthy
and skin-cancer-positive control groups and the highly radiation-sensitive
groups.  Unfortunately, because more finely detailed data on the
radiation sensitivity of the subgroups is not available, it is not
possible here to state whether the individuals in the low sensitivity
group who were clustered with the high sensitivity group had higher
radiation sensitivity than those who did not.
Further scrubbing resulted in residuals that were indistinguishable
from noise (see Methods) and we conclude that only two levels of
structure---corresponding to exposure and high radiation sensitivity---are
present in the data.}

\paragraph{DeSouto Multi-study Benchmark Data}

Having observed the PDM's ability to decouple independent partitions 
in the four-phenotype, three-exposure radiation response data, we
next consider the PDM's ability to articulate disease subtypes.
Because cancers can be molecularly heterogeneous, it is often
important to articulate differences between subtypes---a distinction
that may be more subtle than than the differences caused by
radiation exposure.  Here, we apply the PDM to the suite of 21
Affymetrix data sets previously considered in~\cite{DESO08}.  The
use of these sets is motivated by their diversity and by the 
ability to compare the PDM performance to that of the methods
reported in~\cite{DESO08}.

In~\cite{DESO08}, the authors applied several widely used
clustering algorithms (hierarchical clustering, $k$-means,
finite mixture of Gaussians [FMG], shared nearest-neighbor,
and spectral clustering) to the data using various linkage
and distance metrics as available for each.  In this study,
the number of clusters $k$ was set manually ranging on
$\lceil k_c, \sqrt{n}\rceil$, where $k_c$ is the known number
of sample classes and $n$ is the number of samples; in the 
spectral clustering implementation, $l$ was set equal to the 
chosen $k$.  Note that the PDM differs from basic spectral
clustering as applied in~\cite{DESO08} several crucial ways:
first, $k$ and $l$ are data-driven (thus permitting $k$ that
is smaller than $k_c$, as many dimensions $l$ as are significant
compared to the null model as previously described, and no 
clustering where structure is deemed non-existent compared with
the null model); and second, the successive partitioning 
carried out in the PDM layers can disambiguate mixed clusters.
Importantly, the PDM partitions are obtained without 
relying on prior knowledge of the number of clusters---an
important feature when the data may contain un-identified
disease subtypes.

To illustrate this, we focus on a handful of the benchmark data
sets.  (Full results are provided as supplementary information.)
The partitions are shown in Fig.~\ref{disamb}. In Fig.~\ref{disamb}(a)
and (b), PDM reveals a single layer of three clusters in two versions
of the Golub-1999 leukemia data~\cite{GOLU99}. The two data sets
as provided contained identical gene expression measurements and
differed only in the sample status labels, with Golub-1999-v1 only
distinguishing AML from ALL, but Golub-1999-v2 further distinguishing
between B- and T-cell ALL.  As can be seen from Fig.~\ref{disamb}(a,b),
the PDM articulates a single layer of three clusters, based on the
gene expression data.  In Fig.~\ref{disamb}(a) (Golub-1999-v1), we see that
the AML samples are segregated into cluster 1, while the 
ALL samples are divided amongst clusters 2 and 3; that is, the
PDM partition indicates that there exists structure, distinct from
noise (as defined through the resampled null model), that 
distinguishes the ALL samples as two subtypes.  If we repeat this
analysis with Golub-1999-v2, we obtain the partitions shown in 
Fig.~\ref{disamb}(b).  Since the actual gene expression data
is identical, the PDM partitioning of samples is the same; however, 
we now can see that the division of the ALL samples between
clusters 2 and 3 corresponds to the B- and T-cell subtypes.
One can readily---particularly in the context of cancers---situations
in which unknown sample subclasees exist that could be detected
via PDM (as in Fig.~\ref{disamb}(a)); at the same time, the PDM's
comparison to the resampled null model prevents artificial partitions
of the data.

In Figures~\ref{disamb}(c) and (d), we see how the first layer of
clustering is refined in the second layer; for example, in
Fig.~\ref{disamb}(c), the E2A-PBX1 and T-ALL leukemias are distinguished
in the first layer, while the second serves to separate the MLL and
majority of the TEL-AML subtypes from the mixture of B-cell ALLs
in the first cluster of layer 1. As in Figs.~\ref{disamb}(a) and
(b), the PDM identifies clusters of subtypes that may not be known
a priori (cf results for Yeoh-2002-v1 in Supplement, for which all
the B-cell ALLs had the same class label but were partitioned, as
in Fig.~\ref{disamb}(c), by several subtypes).  In Fig.~\ref{disamb}(d),
second layer cluster assignment in Fig.~\ref{disamb}(d) distinguishes
the ovarian (OV) and kidney (KI) samples from the others in the
mixed cluster 2 in the first layer.

Results for the complete set of Affymetrix benchmark data are given
as Supplementary Information.  A $t$-test comparison of adjusted
Rand indices obtained from the PDM suggests that it is comparable
to those obtained with the best method, FMG, in~\cite{DESO08}.
However, it is important to note that this is achieved by the PDM
in an entirely unsupervised way (in contrast to the heuristic
approach used to select $k$ and $l$ in~\cite{DESO08}), a considerable
advantage.
%
We also note that the PDM performance remained high regardless of
the distance metric used (see supplement), and we did not observe
the large decrease in accuracy noted by \cite{DESO08} when using a Euclidean
metric in spectral clustering.  We attribute this largely to the 
aforemented improvements (multiple layers; data-driven $k$ and $l$
parameterization) of PDM over standard spectral clustering.

\subsection*{Pathway-PDM Analysis}

The above applications of the PDM illustrate its ability to 
detect clusters of samples with common exposures and phenotypes
based on genome-wide expression patterns, without advance
knowledge of the number of sample categories.  However,
it is often of greater interest to identify a set of 
genes that govern the distinction between samples.
Pathway-based application of the PDM permits this by
systematically subsetting the genes in known pathways (here, based on 
KEGG~\cite{KEGG} annotations), and partitioning the samples.
Pathways yielding cluster assignments that correspond to sample
characteristics can then be inferred to be associated with that
characteristic.

We applied pathway-PDM as described above to the radiation response
data from~\cite{REIG04}, testing the clustering results obtained
for inhomogeneity with respect to the phenotype ($\chi^2$ test).
Because some pathways contain a fairly large number of probes, it
is reasonable to ask whether the pathways that permitted clusterings
corresponding to tumor status were simply sampling the overall gene
expression space.  In order to assess this, we also constructed
artificial pathways of the same size as each real pathway by randomly
selecting the appropriate number of probes, and recomputing the
clustering and $\chi^2$ $p$-value as described above.  1000 such
random pathways were created for each unique pathway length, and
the fraction $f_{\mathrm{rand}}$ of 
pathways that yielded a $\chi^2$ $p$-value smaller than that observed
in the ``true'' pathway is used as an additional measure of the
pathway significance.  Six pathways distinguished the radiation-sensitive
samples with $f_{\mathrm{rand}}<0.05$ as shown in Fig.~\ref{radpath};
several also articulated exposure-associated partitions in addition
to the phenotype-associated partition.  Interestingly,
all of the high-scoring pathways separated the high-RS case samples
samples, but did not subdivide the three control sample classes;
this finding, as well as the exposure-independent clustering
assignments in several pathways in Fig.~\ref{radpath}, suggests
that there are systematic gene expression differences between the
radiation-sensitive patients and all others.  Several other pathways
(see Supplemental Information) yield exposure-associated partitions
without distinguishing between phenotypes; unsurprisingly, these
are the cell cycle, p53 signaling, base excision repair, purine
metabolism, MAP kinase, and apoptosis pathways.

To further illustrate Pathway-PDM, we apply it to the Singh prostate
gene expression data~\cite{SING02} (the heavily-filtered sets from
\cite{DESO08} have too few remaining probes to meaningfully subset by pathway).
First, we observe that in the complete gene expression space, the clustering
of samples corresponds to the tumor status in the second PDM layer
(see Supplemental Information).  This is consistent with the molecular
heterogeneity of prostate cancer, and suggests that the first layer 
describes individual variation that is scrubbed out and then revealed 
in the second layer.  Next, we apply pathway-PDM as described above,
testing each layer of clustering for inhomogeneity
with respect to the known tumor/normal labels ($\chi^2$ test).
%

%

Of the 203 pathways considered, those that yielded significant
$f_{\mathrm{rand}}$ in any layer of clustering is given in
Table~\ref{prostab}.  No pathway pathway yielded more than
two layers of structure.  A total of 29 of 203 pathways 
exhibited significant clustering inhomogeneity in any layer;
amongst the significant pathways, the misclassification
rate---the fraction of tumor samples that are placed in a cluster that is 
majority non-tumor and vice-versa---is approximately 20\%.  Plots of
the six most discriminative pathways in layers 1 and 2 are given in 
Figure~\ref{prospath}.

A number of known prostate cancer related pathways appear at the
top of this list.  The urea acid cycle pathway, prion disease
pathway, and bile acid synthesis pathways have previously been noted
in relationship to prostate cancer~\cite{MANO06}.  The coagulation
cascade is known to be involved in tumorigenesis through its role
in angiogenesis~\cite{RICK03}, and portions of this pathway have
been implicated in prostate metastasis~\cite{KLEZ04}.  Cytochrome
P450, which is part of the inflammatory response, has been implicated
in many cancers~\cite{AGUN04}, including prostate~\cite{MURA01},
with the additional finding that it may play a role in estrogen
metabolism (critical to certain prostate cancers)~\cite{TSUC05}.
Many amino-acid metabolism pathways (a hallmark of proliferating
cells) and known cancer-associated signaling pathways (Jak-STAT,
Wnt) are also identified.

Because pathway-PDM does not rely upon single-gene associations and
employs a ``scrubbing'' step to reveal progressively finer
relationships, we expect that we will be able to identify pathways
missed by other methods.  It is of interest to compare the results
obtained by pathway-PDM to those obtained by other pathway analysis
techniques.  In~\cite{MANO06}, the authors applied several established
pathway analyses (Fisher's test, GSEA, and the Global Test) to a
suite of three prostate cancer gene expression data sets, including
the Singh data considered here.  Fifty-five KEGG pathways were
identified in at least one data set by at least one method~\cite{MANO06},
but with poor concordance:
15 of the these were found solely in the Singh data, and 13 were found
in both the Singh data and at least one of the other two data sets
(Welsh~\cite{WELS02}, Ernst~\cite{ERNS02}) using any method.
A comparison of the pathway-PDM identified pathways to those 
reported in~\cite{MANO06} is given by the final column of Table~\ref{prostab},
which lists the data sets that
yielded significance by any method (Fisher's test, GSEA, and the
Global Test) reported in~\cite{MANO06}.  
Of the 29 pathway-PDM identified pathways, 16 had been identified
by~\cite{MANO06} in either the Welsh or Ernst data (including
7 found by other methods in the Singh data by~\cite{MANO06}).
%
%
The PDM-identified pathways show improved concordance
with the pathways identified in~\cite{MANO06};
while only 13 of the 40
pathways identified in the Welsh or Ernst data were corroborated by the Singh
data using any method in~\cite{MANO06}, the addition of the pathway-PDM
Singh results brings this to 22/40. 
Of the 13 pathways newly
introduced in Table~\ref{prostab}, several are already known
to play a role in prostate cancer but were not detected
using the methods in~\cite{MANO06} (such as cytochrome P450,
complement and coagulation cascades, and Jak-STAT signalling);
several also constitute entries in KEGG that were either not
present at the time that~\cite{MANO06} was published or have
had over 30\% of genes added/removed, making them incomparable
to the KEGG annotations used in~\cite{MANO06}.
This improved concordance supports the inferred role the 
PDM-identified pathways in prostate cancer, and suggests that
the pathway-PDM is able to detect pathway-based gene expression
patterns missed by other methods as applied to the Singh data.

\comment{
A number of cancer-related pathways appear at the top of this list.
The coagulation cascade is known to be involved in tumorigenesis
through its role in angiogenesis~\cite{RICK03}, and portions of
this pathway have been implicated in prostate metastasis~\cite{KLEZ04}.
Cytochrome P450, which is part of the inflammatory response, has
been implicated in many cancers~\cite{AGUN04}, including
prostate~\cite{MURA01}, with the additional finding that it may
play a role in estrogen metabolism (critical to certain prostate
cancers)~\cite{TSUC05}.  Amino-acid
metabolism (a hallmark of proliferating cells) and known cancer-associated
signaling pathways (Jak-STAT, Wnt).

androgen
and estrogen metabolism, DNA replication, other cancers (melanoma)
and inflammatory responses (arachidonic acid), and the tumor-suppressor
p53 signaling mechanism are also notably present as having
pathway-wide differences that permit clustering of tumor samples.}

\comment{
Because prostate cancer is known to be histologically
diverse~\cite{SING02}, we believe we will find
phenotype-related structure on the second level of the PDM in pathways
for which the first layer was dominated by non-cancer biological
differences.   To investigate this, we carried out the scrubbing
and clustering steps of the PDM on each of the pathways, with
highly-significant results given in Table~\ref{pros2}.  As with the
significant first-layer significant pathways, the misclassification
rate---the fraction of tumor samples that are placed in a cluster
that is majority non-tumor and vice-versa---is approximately 20\%.
Plots of the top three pathways are given in Figure~\ref{prosB}.
Once again, pathways related to the inflammatory response, cell growth,
and cancers---including the prostate cancer pathway---are  present.

It is notable that a larger fraction of pathways met the significance
threshold for class prediction in the second layer than in the first
(Table~\ref{proscount}).  This suggests that biologically-relevant
differences between tumor and non-tumor cells are likely to exist
at a finer scale than that detected in the first PDM layer, and supports
our assertion above that structure in the first layer is a result of the 
histological diversity of prostate tumors and corresponds to biological
traits that are independent of tumor status.

Further scrubbing and clustering iterations beyond the
second layer  resulted in more partition failures (that is, after
scrubbing fewer pathways had structure distinguishable from noise)
and fewer pathways met the significance threshold for class prediction
in the higher layers.  A summary of the number of pathways with
structure distinguishable from noise and structure corresponding
to tumor status is given in Table~\ref{proscount}.
}

\comment{

\subsection*{Incorporating Yu prostate data}

The PDM may also be used to ``scrub out'' differences that are due to 
different microarray conditions, which we illustrate using the Singh~\cite{SING02} and
Yu~\cite{YU04} data sets together.  We recall here the fact that the samples were
from different study populations, hybridized to slightly different arrays, and 
normalized separately using different algorithms.  We combine these disparate
data sets into a single matrix, retaining the genes assayed by both, and
applied the PDM.  Results are shown in Tables~\ref{prosboth1}-\ref{prosboth2} and 
Figure~\ref{prosboth}. 

As anticipated, the first layer of structure corresponds to the
study, with the first two clusters (three clusters were automatically
chosen) corresponding to the Singh data and the third corresponding
to Yu.  After scrubbing this variation from the data and clustering
the residuals in the second PDM layer, we are left with structure
that correlates strongly with phenotype: the first cluster has all
of the normal and majority of the stromal samples from both studies,
and the second cluster has all of the metastatic and the majority
of tumor  samples from the combined studies.  This strongly suggests
that there exist genome-wide patterns of expression that correlate
with prostate cancer phenotypes, and suggests comparability between
disparate studies; that is, after scrubbing we find that the normal
cells and cancer cells cluster together (with 63\% sensitivity and
86\% specificity), regardless of the data source.  The same analysis
may be carried out on a pathway-by-pathway basis, as was done in the
Singh data, first scrubbing the variation from the disparate studies and
then finding pathways that permit classification of tumor and normal
samples regardless of source; pathways with high specificity 
and sensitivity are given in Table~\ref{prosbothbypath}.

To mimic a situation in which we have a classifier that
works in the Singh data and a single new unknown sample for which
we have a measure of gene expression that may not be comparable,
in absolute terms, to the original data (e.g., in the case where the
new sample is assayed by another chip or PCR), 
we extend the pathway-by-pathway clustering of the Singh 
data to perform class prediction in the Yu data set.
In order to find pathways that would permit identification of the phenotype 
of a new, unmatched sample, we incorporated each sample from the
Yu data into the clustering as described in the methods, using the
top ten pathways from Table~\ref{pros1} (i.e., those that permitted
highly accurate class predictions of the Singh data).  
Three pathways---metabolism
of xenobiotics by cytochrome P450, tyrosine metabolism, and urea
cycle and metabolism of amino groups---are able to distinguish
phenotypes in the Yu data in spite of systematic differences in
the data sets, as shown in Tables~\ref{luo1}-\ref{luo3}.  Using the
cytochrome P450 pathway (Table~\ref{luo1}), prostate tumor and
metastatic cells are identified with 70\% sensitivity, but low
(50\%) specificity.  The tyrosine metabolism pathway (Table~\ref{luo2})
does substantially better, yielding a 90\% sensitive identification
of prostate tumor and metastatic cells, with 55\% specificity for
normal cells; stromal cells here are mistakenly identified as tumor
53\% of the time, corresponding to the finding (in~\cite{YU04} and
elsewhere) that stromal tissue often presents abnormalities consistent
with a tumor ``field effect.'' The urea cycle and metabolism of
amino groups pathway (Table~\ref{luo3}) finds aggressive tumor cells
(those in metastatic tissue) with 80\% sensitivity and 72\% specificity
for non-tumor and stromal tissue, while non-metastatic tumor cells
are often classified as non-tumor.

While imperfect, the accuracy of these results is surprising and
highly encouraging.  We expected that differences in the study
populations, microarray platform, and normalization would dominate
gene expression differences (effectively adding a large amount of
systematic noise that would be avoidable in a more stringent setting,
but likely to be uncontrollably present in a clinical application).
Indeed, seven of the ten pathways tested do not permit classification
of the Yu samples by clustering with the Singh data; the differences
between the Yu and Singh samples are such that the vast majority
of the Yu samples get categorized with a single group of Singh
(i.e., all identified as tumor or all identified as non-tumor).
Finding pathways such as those in Tables~\ref{luo1}-\ref{luo3} that
permit class predictions in the presence of this noise is an important
step in ensuring that the pathway-based findings are generalizable
enough to be of clinical use.
}

\section*{Discussion}

\comment{
\begin{verbatim}
* better than other methods
	- no reliance on single-gene statistics
	- nonconvex boundaries
	- no heuristics: data driven (null model) dim. reduction, 
		cluster number estimation
	- restate comparison to SPACC QUBIC GPC -- cf intro

* good for identifying teasing out variation/ IDing subtypes:
	- in radiation data, high clustering accuracy (better than
		original analysis)
	- articulation of independent partitions in the data that
		reflect the 4x3 study design and discriminate UV, IR exposed 
		high-RS cells from mock exposed high-RS cells AND UV, IR i
		exposed control cells
	- in deSouto data, automatic detection of subtypes in
		unsupervised manner
	- as good or better results than deSouto over all, without
		forcing cluster number

* pathway-PDM for finding pathways
	- expected players in radiation response data
	- prostate pathways: many previously reported
	- better concordance with pathway results from other
		data sets (Manoli) -- pathway-PDM getting at 
		underlying "truths" missed with the other methods
		in the Singh data?
	- detection of several additional, biologically plausible
		pathways not reported in Manoli's analysis
	- need not confine the chi2 test to known class labels;
		instead, can use cluster labels from full gene expr
		PDM run (which may partition UNKNOWN subtypes) and
		look for pathways driving the genome-wide distinction.

\end{verbatim}
}

We have presented here a new application of the Partition Decoupling
Method~\cite{LEIB08,LEIB09} to gene expression profiling data,
demonstrating how it can be used to identify multi-scale relationships
amongst samples using both the entire gene expression profiles
and biologically-relevant gene subsets (pathways).  By comparing the
unsupervised groupings of samples to their phenotype, we use the PDM to
infer pathways that play a role in disease.

The PDM has a number of features that make it preferable to existing
microarray analysis techniques.  First, the use of spectral clustering
allows identification of clusters that are not necessarily separable
by linear surfaces, permitting one to identify complex relationships
between samples.  Importantly, this means that clusters of samples
can be identified even in situations where the genes do not exhibit
differential expression (ie, when they are not linearly separable);
this is particularly useful when examining gene expression profiles
of complex diseases, where single-gene etiologies are rare. 
We observe the benefit of this feature in the example of 
Fig.~\ref{spellcyc}, where the two separate yeast cell groups
could not be separated using $k$-means clustering but could be
correctly clustered using spectral clustering, and we note that
the oscillatory nature of many genes~\cite{PTIT07}
makes detecting such patterns crucial.

Second, the PDM employs not only a low-dimensional embedding of the 
feature space, thus reducing noise (an important consideration when
dealing with noisy microarray data), but also the optimal dimensionality
and number of clusters is data-driven rather than heuristically set.
This makes the PDM an entirely unsupervised method.  Because those
parameters are obtained with reference to a resampled null model,
the PDM prevents samples from being clustered when the relationships
amongst them are indistinguishable from noise.  We observed the
benefit of this feature in the radiation response data~\cite{REIG04}
shown in Fig.~\ref{pdmfig}, where two (as opposed to four)
phenotype-related clusters were articulated by the PDM: the first
corresponding to the high-RS cases, and the second corresponding
to a combination of the three control groups.

Third, the independent ``layers'' of clusters (decoupled partitions)
obtained in PDM provide a natural means of teasing out variation
due to experimental conditions, phenotypes, molecular subtypes, and
non-clinically relevant heterogeneity.  We observed this in the
radiation response data~\cite{REIG04}, where the PDM identified the
exposure groups with 100\% accuracy in the first layer (Fig.~\ref{pdmfig}
and Table~\ref{tab1}) followed by highly accurate classification
of the high-RS samples in the second layer (Fig.~\ref{pdmfig} and
Table~\ref{tab4}).  The improved sensitivity to classify high-RS
samples over linear methods (83\% vs. the 64\% reported using SAM
in~\cite{REIG04}) suggests that there may exist strong patterns,
previously undetected, of gene expression that correlate with
radiation exposure and cell type.  This was also observed in the
benchmark data sets~\cite{DESO08}, shown in Fig.~\ref{disamb} and
Supplement, where the PDM automatically detected subtypes in an
unsupervised manner without forcing the cluster number.  The results
from PDM in the radiation response data and benchmark data sets
were as or more accurate than those reported using other algorithms
in~\cite{REIG04,DESO08}, were obtained without reliance upon
heuristics, and reflect statistically significant (with reference
to the resampled null model) relationships between samples in the
data.

The accuracy of the PDM can used, in the context of gene subsets
defined by pathways, to identify mechanisms that permit the
partitioning of phenotypes.  In Pathway-PDM, we subset the genes
by pathway, apply the PDM, and then test whether the PDM cluster
assignments reflect the known sample classes.  Pathways that permit
accurate partitioning by sample class contain genes with expression
patterns that distinguish the classes, and may be inferred to play
a role in the biological characteristics that distinguish the
classes.  To illustrate this, we applied Pathway-PDM to both the
radiation response data~\cite{REIG04} and a prostate cancer data
set~\cite{SING02}.  In the radiation response data~\cite{REIG04},
we identified pathways that partitioned the samples by phenotype
and both by phenotype and exposure (Fig.~\ref{radpath}) as well as
pathways that only partitioned the samples by exposure without
distinguishing the phenotypes (Supplement).  In the prostate cancer
data~\cite{SING02}, we identified 29 pathways that partitioned the
samples by tumor/normal status (Table~\ref{prostab}).  Of these,
15 revealed the significant tumor/normal partition in the second
layer rather than the first (as did the full-genome PDM, Supplement),
and 13 of the 14 pathways with significant tumor/normal partitions
in the first layer contained additional non-noise structure in the
second; prostate cancer is known to be molecularly diverse~\cite{SING02},
and these partitions may reflect unidentified subcategories of
cancer or some other heterogeneity amongst the patients.  By applying
Pathway-PDM to the Singh data, we were able to improve upon the
concordance reported using pathway-based analyses in~\cite{MANO06},
suggesting that the accuracy of PDM may permit the identification
of significant pathways, via Pathway-PDM, that are missed by other
methods.  

While our application of Pathway-PDM was such that PDM clusters
from each pathway were compared against known sample class labels,
one can just as easily compare them to labels from the cluster
assignment from full-genome PDM.  Hence, for example, in a situation
such as the Golub-1999-v1 data shown in Fig.~\ref{disamb}(a), one
could use the 3-cluster assignment, rather than the 2-class sample
labels, to find the pathways that permit the separation of cluster-2
ALLs from the cluster-3 ALLs.  In a case like this, where full-genome
PDM reveals subtypes are not already known, applying Pathway-PDM
may help identify the molecular mechanisms driving the subtype.

Despite these clear benefits, the PDM as applied here has a drawback:
specifically, the low-dimensional nonlinear embedding of the data
that makes spectral clustering and the PDM powerful also
complicates the biological interpretation of the findings (in much
the same way that clustering in principal component space might).
Pathway-PDM serves to address this issue by leveraging expert
knowledge to identify mechanisms associated with the phenotypes.
Additionally, the nature of the embedding, which relies upon the
geometric structure of all the samples, makes the classification
of an entirely new sample challenging.  These issues might be
addressed in several ways: experimentally, by investigation of the
Pathway-PDM identified pathways (possibly after further subsetting
the genes to subsets of the pathway) to yield a better biological
understanding of the dynamics of the system that were ``snapshot''
in the gene expression data; statistically, by modeling outcome the
pathway genes using an approach such as~\cite{BAKE10} that explicitly
accounts for oscillatory patterns (as seen in Fig.~\ref{spellcyc})
or such as~\cite{HUDS09} that accounts for the interaction structure
of the pathway; or geometrically, by implementing an out-of-sample
extension for the embedding as described in~\cite{BENG04a,BENG04}
that would allow a new sample to be classified against the PDM
results of the known samples.

In sum, our findings illustrate the utility of the PDM in gene
expression analysis and establish a new technique for pathway-based
analysis of gene expression data that is able to articulate phenotype
distinctions that arise from systems-level (rather than single-gene)
differences.  We expect this approach to be of use in future
analysis of microarray data as a complement to existing linear
techniques.

\comment{
We have presented here a new application of the Partition Decoupling
Method~\cite{LEIB08,LEIB09} to gene expression profiling data,
demonstrating how it can be used to identify multi-scale relationships
amongst samples using both the entire gene expression profiles
and biologically-relevant gene subsets (pathways).  By comparing the
unsupervised groupings of samples to their phenotype, we use the PDM to
infer pathways that play a role in disease.

The PDM has a number of features that make it preferable to existing
microarray analysis techniques.  The use of spectral clustering
allows identification of clusters that are not necessarily separable
by linear surfaces, permitting one to identify complex relationships
between samples.  Importantly, this means that clusters of samples
can be identified even in situations where the genes do not exhibit
differential expression (ie, when they are not linearly separable);
this is particularly useful when examining gene expression profiles
of complex diseases, where single-gene etiologies are rare.  The PDM
uses a low-dimensional embedding of the feature space, an important
consideration when dealing with noisy microarray data.  Because the
data itself is used to determine both the optimal number of clusters
and the optimal dimensionality in which the feature space is
represented, the PDM provides an entirely unsupervised method for
classification without relying upon heuristics.  By scrubbing the
data and repeating the clustering on the residuals, finer relationships
may be revealed.

To illustrate its utility, we applied the PDM both across the complete
gene expression profile and on a pathway-by-pathway basis in three
gene expression data sets: one from a radiation response
study~\cite{REIG04}, and two from prostate cancer
studies~\cite{SING02,YU04}.
The results of the PDM applied to the radiation response data permit
us to conclude that two layers of structure, corresponding to
radiation exposure and radiation sensitivity, are present in the
gene expression data.   While radiation sensitivity is weakly
discernible using traditional clustering methods on the original
data, we find that scrubbing out the exposure-related structure
reveals a much cleaner clustering of cell-type using spectral
clustering.  Notably, the PDM not only identifies exposure groups with
100\% accuracy (Fig.~\ref{pdmfig}(a) and Table~\ref{tab1}), but also
permits us to improve considerably the classification of
radiation-sensitive cells to 83\% from the 64\% sensitivity reported
in~\cite{REIG04} (Fig.~\ref{pdmfig}(b) and Table~\ref{tab4}).  This
is a considerable improvement, and it suggests that there exist
strong patterns, previously undetected, of gene expression that
correlate with radiation exposure and cell type.

In the Singh prostate data, we demonstrated how the PDM may be used to
find pathways that permit classification of tumor and non-tumor
tissue.  These pathways contain genes that exhibit patterns amongst
tumor samples that distinguish them from non-tumor tissue, despite
the diversity of prostate tumors.  Pathways discovered as significant
in pathway-PDM analyses of gene expression microarray data are
likely to be relevant to disease progression and may be followed
up by functional studies that target specific systems
(Tables~\ref{pros1},~\ref{pros2}; Figs.~\ref{prosA},~\ref{prosB}).
By using the PDM rather than looking for an overabundance of differentially
expressed genes within a pathway, pathways with patterns of gene
expression that do not manifest as differential single-gene expression
(such as the toy example Fig.~\ref{twocircs} described in the
methods) can be revealed.

Following the observation that the PDM can tease out multi-scale structure
in the radiation response data, we showed how the PDM can be used on
combined data sets. Instead of normalizing the combined data, the PDM
was used first to extract variation due to the disparate data sets;
the residuals may then be examined by a second layer of clustering,
either using the whole gene expression profile (Fig.~\ref{prosboth};
Tables~\ref{prosboth1},~\ref{prosboth2}) or individual pathways
(Table~\ref{prosbothbypath}).  The residual data from the first
clustering and scrubbing step may also be analyzed in the usual way
to find genes with differential expression.   Because the clustering
step that precedes the scrubbing permits clusters with nonlinear
separations, scrubbing the dataset-related variation using the PDM
permits the extraction of signals that would not be found using, for example,
linear regression with the study as one of the independent
variables.  The PDM may thus be used to combine disparate studies and
potentially improve the comparability of microarray results.

We also showed how several of the pathways identified as relevant
in the Singh data could be used to classify a new sample (taken
from the Yu data) without the requirement that the gene expression
be measured on the same platform or that the new sample's data be
normalized to the data which define the clusters
(Tables~\ref{luo1}-\ref{luo3}).  Our findings here suggest two
things: first, that these pathways exhibit differences in gene
expression that are generalizable beyond the Singh study, making
them of interest in further investigation; and second, that it is
possible, using the PDM, to devise a classifier that will be robust to
the measurement platform.

In sum, our findings illustrate the utility of the PDM in gene expression
analysis and establish a new technique for pathway-based analysis of
gene expression data that is able to articulate phenotype distinctions
that arise from systems-level (rather than single-gene) differences.
We expect this approach to be of great use in future analysis of
microarray data as a companion to existing linear techniques.
}

\section*{Acknowledgments}
RB would like to thank Sean Brocklebank (University of Edinburgh)
for many fruitful discussions.  This work was made possible by the
Santa Fe Institute Complex Systems Summer School (2009).  RB is
supported by the Cancer Prevention Fellowship Program and a Cancer
Research Training Award, National Cancer Institute, NIH.

\bibliographystyle{plos2009}
\bibliography{/h1/braunr/Papers/tex-repo/journals_short,/h1/braunr/Papers/tex-repo/rb-nci}
\clearpage
\setlength{\tabcolsep}{1em}
\renewcommand{\thetable}{\arabic{table}}

\begin{table}
\begin{center}
\begin{tabular}{ll}
\hline
&\textbf{Spectral Clustering Algorithm}\\
\hline
1.& Compute the correlation $\rho_{i,j}$ between all pairs of $n$ data points $i$ and $j$.\\
2.& Form the affinity matrix $\mat{S}\in \mathbb{R}^{n\times n}$ defined by $s_{i,j} = \exp \big[-\sin^2\big(\arccos(\rho_{i,j})\big)/\sigma^2\big]$,  where \\ & $\sigma$ is a scaling parameter ($\sigma=1$ in the reported results).\\
3.& Define $\mat{D}$ to be the diagonal matrix whose $(i,i)$ element is the column sums of $\mat{S}$.\\
4.& Define the Laplacian $\mat{L} = \mat{I} - \mat{D}^{-1/2}\mat{S}\mat{D}^{-1/2}$.\\
5.& Find the eigenvectors $\{v_0, v_1, v_2, \dots, v_{n-1}\}$ with corresponding eigenvalues \\
& $0 \leq \lambda_1 \leq \lambda_2 \leq \dots \leq \lambda_{n-1}$ of $\mat{L}$. \\
6.& Determine from the eigendecomposition the optimal dimensionality $l$ and natural\\
&  number of clusters $k$ (see text).\\
7.& Construct the embedded data by using the first $l$ eigenvectors to provide coordinates \\ & for the data (i.e., sample $i$ is assigned to the point in the Laplacian eigenspace with \\ & coordinates given by the $i$th entries of each of the first $l$ eigenvectors, similar to PCA).\\ 
8.& Using $k$-means, cluster the $l$-dimensional embedded data into $k$ clusters.\\
\hline
\end{tabular}
\end{center}
\caption{ \label{specclus}Procedure for Spectral Clustering.}
\end{table}

\clearpage

\begin{table}
\begin{center}
\begin{tabular}{rr|rrr}
\hline
&& \multicolumn{3}{c}{Cluster} \\
&&  1 & 2 & 3\\
\hline
\multirow{3}{*}{Treatment} & Mock & 57 & 0 & 0\\
& IR & 0 & 57 & 0\\
& UV & 0 & 0 &57\\ 
\hline
\end{tabular}
\end{center}
\caption{\label{tab1} Spectral clustering of expression data versus exposure; exposure categories are reproduced exactly.}
\end{table}

\begin{table}
\begin{center}
\begin{tabular}{rr|rrr}
\hline
&& \multicolumn{3}{c}{Cluster} \\
&&  1 & 2 & 3\\
\hline
\multirow{3}{*}{Treatment} & Mock & 36 & 15 & 6\\
& IR & 36 & 15 & 6\\
& UV & 3 & 14 & 40\\
\hline
\end{tabular}
\end{center}
\caption{\label{tab2} $k$-means clustering of expression data versus exposure using $k=3$.}
\end{table}

\begin{table}
\begin{center}
\begin{tabular}{rr|rrrr}
\hline
&& \multicolumn{4}{c}{Cluster} \\
&&  1 & 2 & 3 & 4\\
\hline
\multirow{4}{*}{Cell type} & Healthy & 19 & 18 & 8 & 0\\
& Skin cancer  & 8 & 23 & 14 & 0\\
& Low radiation sensitivity  & 13 & 11 & 8 & 7\\
& High radiation sensitivity& 6 & 1 & 9 & 26\\
\hline
\end{tabular}
\end{center}
\caption{\label{tab3} $k$-means clustering of expression data versus cell type using $k=4$.}
\end{table}

\begin{table}
\begin{center}
\begin{tabular}{rr|rr}
\hline
&& \multicolumn{2}{c}{Cluster} \\
&&  1 & 2 \\
\hline
\multirow{4}{*}{Cell type} & Healthy & 45 & 0 \\
& Skin cancer  & 45 & 0 \\
& Low radiation sensitivity& 28 & 11\\
& High radiation sensitivity& 7 & 35\\
\hline
\end{tabular}
\end{center}
\caption{\label{tab4} Spectral clustering of exposure data with exposure-correlated clusters scrubbed out, versus cell type.}
\end{table}

\clearpage
\begin{table}[ht]
\begin{center}
{\footnotesize{
\begin{tabular}{@{~}r@{~~~}l|r|rr|rr|l@{~}}
&&&\multicolumn{2}{c|}{Layer 1} &\multicolumn{2}{c|}{Layer 2}& \\
&KEGG Pathway & $L_p$ & $p$ $(\chi^2)$ & $f_\mathrm{rand}$ & $p$ $(\chi^2)$ & $f_\mathrm{rand}$& In~\cite{MANO06}~?\\
  \hline

	00220 & Urea cycle \& metabolism of amino groups & 33 & 1.14e-13 & $<0.001$ & 7.10e-01 & 0.940 & \cite{SING02,WELS02,ERNS02}\\ 
  00980 & Metab. of xenobiotics by cytochrome P450 & 72 & 3.97e-13 & 0.001 &  &  & --- \\ 
  00640 & Propanoate metabolism & 31 & 7.78e-12 & 0.003 & 9.78e-01 & 0.995& \cite{WELS02,ERNS02}\\ 
  04610 & Complement and coagulation cascades & 75 & 9.21e-12 & 0.008 & 2.47e-02 & 0.371& \\ 
  00120 & Bile acid biosynthesis & 32 & 1.29e-01 & 0.699 & 1.15e-11 & 0.003& \cite{SING02,WELS02}\\ 
  05060 & Prion disease & 18 & 5.18e-02 & 0.527 & 2.20e-11 & 0.003& \cite{SING02,WELS02,ERNS02}\\ 
  00380 & Tryptophan metabolism & 50 & 3.84e-11 & 0.008 & 5.52e-01 & 0.894& \cite{ERNS02}\\ 
  00480 & Glutathione metabolism & 48 & 4.80e-11 & 0.008 & 8.37e-01 & 0.955& \cite{SING02,WELS02,ERNS02}\\ 
  04310 & Wnt signaling pathway & 191 & 5.38e-11 & 0.017 & 5.47e-01 & 0.916& \cite{WELS02}\\ 
  00983 & Drug metabolism - other enzymes & 52 & 5.08e-10 & 0.024 & 8.60e-01 & 0.966&  --- \\ 
  04630 & Jak-STAT signaling pathway & 205 & 1.65e-01 & 0.826 & 8.41e-10 & 0.025& \\ 
  00053 & Ascorbate and aldarate metabolism &  8 & 3.32e-02 & 0.462 & 7.67e-09 & 0.008& \cite{ERNS02}\\ 
  00350 & Tyrosine metabolism & 45 & 1.32e-02 & 0.359 & 2.80e-08 & 0.040& \cite{SING02,WELS02}\\ 
  00641 & 3-Chloroacrylic acid degradation & 16 & 5.23e-08 & 0.016 & 6.89e-01 & 0.893& \\ 
  00960 & Alkaloid biosynthesis II &  8 & 7.13e-02 & 0.558 & 8.23e-08 & 0.016& \cite{SING02}\\ 
  00410 & beta-Alanine metabolism & 25 & 9.24e-08 & 0.016 & 1.60e-01 & 0.673& \cite{ERNS02}\\ 
  00650 & Butanoate metabolism & 37 & 9.39e-02 & 0.645 & 1.50e-07 & 0.014& \\ 
  00260 & Glycine, serine \& threonine metabolism & 36 & 9.56e-02 & 0.645 & 1.78e-07 & 0.014& \cite{WELS02,ERNS02}\\ 
  00600 & Glycosphingolipid metabolism & 32 & 7.84e-02 & 0.615 & 3.08e-07 & 0.016& \cite{SING02}\\
  00030 & Pentose phosphate pathway & 21 & 3.59e-07 & 0.022 & 2.80e-01 & 0.755& \cite{WELS02,ERNS02}\\ 
  00062 & Fatty acid elongation in mitochondria & 11 & 1.68e-01 & 0.684 & 3.67e-07 & 0.022& \cite{SING02,WELS02}\\ 
  00272 & Cysteine metabolism & 10 & 6.01e-07 & 0.025 & 7.52e-02 & 0.574& --- \\ 
  00340 & Histidine metabolism & 27 & 3.94e-02 & 0.477 & 1.42e-06 & 0.022& \cite{ERNS02}\\ 
  00720 & Reductive carboxylate cycle &  9 & 7.62e-02 & 0.574 & 1.51e-06 & 0.025& \cite{SING02}\\ 
  00565 & Ether lipid metabolism & 23 & 4.07e-06 & 0.036 & 8.43e-01 & 0.948&  --- \\ 
  01032 & Glycan structures - degradation & 39 & 8.17e-01 & 0.957 & 4.62e-06 & 0.038& \\ 
  00360 & Phenylalanine metabolism & 19 & 2.32e-02 & 0.376 & 6.26e-06 & 0.044& \cite{WELS02,ERNS02}\\ 
  00040 & Pentose and glucuronate interconversions & 17 & 7.75e-06 & 0.047 & 4.98e-01 & 0.843& \cite{SING02}\\ 
  00051 & Fructose and mannose metabolism & 35 & 4.49e-03 & 0.211 & 7.99e-06 & 0.043& \cite{SING02,WELS02}\\ 

\end{tabular} 
}}
\caption{ \label{prostab} Pathways with cluster assignment articulating tumor versus normal status in at least one PDM layer for the Singh prostate data.  The $L_p$ column lists the size of the pathway. $\chi^2$ test $p$-values for tumor status versus cluster assignment in PDM layer 1 and layer 2 are given.  The $f_\mathrm{rand}$ columns show the fraction of randomly-generated pathways with smaller $\chi^2$ $p$-values in either PDM layer. The final column lists the data sets for which~\cite{MANO06} identified the pathway as significant (\cite{SING02}, Singh; \cite{WELS02}, Welsh; \cite{ERNS02}, Ernst;  a dash indicates pathways with significant revisions ($>$30\% of genes added or removed) in KEGG between this analysis and the time of \cite{MANO06} publication).}
\end{center} 
\end{table}



\newcommand{\prosAacap}{}
\newcommand{\prosAbcap}{}
\newcommand{\prosAccap}{}
\newcommand{\prosBacap}{}
\newcommand{\prosBbcap}{}
\newcommand{\prosBccap}{}
\newcommand{\prosbothacap}{}
\newcommand{\prosbothbcap}{}
\newcommand{\pdmradAcap}{}
\newcommand{\pdmradBcap}{}
\newcommand{\pdmradCcap}{}
\newcommand{\disambAcap}{}
\newcommand{\disambBcap}{}
\newcommand{\disambCcap}{}
\newcommand{\disambDcap}{}

\newcommand{\disambcap}{
{PDM results for several benchmark data sets.}
{Points are placed in the grid according to
cluster assignment from layers 1 and 2 (in
(a) and (b) no second layer is present). In (a) and (b) it can be seen
that the PDM identifies three clusters, and that the division of the ALL
samples in (a) corresponds to a subtype difference (ALL-B, ALL-T) in (b).
In (c) and (d), it can be seen that the partitioning of samples in the first layer is refined in the second PDM layer.}
}

\newcommand{\radpathcap}{
{Pathway-PDM results for top pathways in radiation response data.}
{Points are placed in the grid according to
cluster assignment from layers 1 and 2 along for 
pathways with $f_\mathrm{rand}<0.05$.  Exposure is indicated by
shape (``M'', mock; ``U'', UV; ``I'', IR), with phenotypes (healthy, skin cancer, low RS, high RS)
indicated by color. Several pathways (nucleotide excision repair,
Parkinson's disease, and DNA replication) cluster samples 
by exposure in one layer and phenotype in the other, suggesting that these mechanisms
differ between the case and control groups.}}

\newcommand{\prospathcap}{
{Pathway-PDM results for top pathways in the Singh prostate data.}
{Points are placed in the grid according to
cluster assignment from layers 1 and 2; shown are the six
most discriminative pathways.}}

\newcommand{\pdmradcap}{
{PDM results for radiation response data.}
{
Shown in (a) and (b) is a scatter plot of each sample's Fiedler
vector value along with the resulting clustering (indicated by color)
for the first (a) and second (b) PDM layers.  A Gaussian mixture
fit to the density (left panel) of the Fiedler vector is used to 
assess the number of clusters, and the resulting cluster assignment for 
each sample is
indicated by color. Exposure is
indicated by shape (``M'', mock; ``U'', UV; ``I'', IR), with
phenotypes (healthy, skin cancer, radiation insensitive,
radiation sensitive) grouped together along the $x$-axis.
In~(a), it can be
seen that the cluster assignment correlates with exposure, while in~(b),
cluster assignment correlates 
with radiation sensitivity.  In (c), points are placed in the grid according to
cluster assignment from layers 1 and 2 along the $x$ and $y$ axes; it 
can be seen that the UV- and IR- exposed high-sensitivity samples differ
both from the mock-exposed high-sensitivity samples as well as the UV- and IR- exposed control samples.}}

\newcommand{\pdmfigonecap}{{Laplacian matrix eigenvalues (top) and
Fiedler vector values (bottom) for spectral clustering of radiation
exposure data. }  {In the top plot, the resampling-based threshold for
eigenvalue significance is shown in cyan, with smaller eigenvalues
plotted in red.  In the bottom plot, we show each sample's Fiedler
vector value along with the resulting clustering.  A Gaussian mixture
fit to the density (bottom left) of the Fiedler vector indicates
two clusters; the resulting cluster assignment for each sample is
indicated by color. True treatment categories of each sample are
given as shapes: crosses denote mock; circles, UV; triangles, IR.
The four cell types (healthy, skin cancer, radiation insensitive,
radiation sensitive) are separated by vertical lines. It can be
seen that the cluster assignment correlates precisely with the
exposure type, independent of cell type.}}

\newcommand{\pdmfigtwocap}{{Laplacian matrix eigenvalues (top) and
Fiedler vector values (bottom) for spectral clustering of scrubbed
radiation exposure data. } {In the top plot, the resampling-based
threshold for eigenvalue significance is shown in cyan, with smaller
eigenvalues plotted in red.  In the bottom plot, we show each
sample's Fiedler vector value along with the resulting clustering.
A Gaussian mixture fit to the density (bottom left) of the Fieldler
vector indicates three clusters; the resulting cluster assignment
for each sample is indicated by color. True treatment categories
of each sample are given as shapes: crosses denote mock; circles,
UV; triangles, IR. The four cell types (healthy, skin cancer,
radiation insensitive, radiation sensitive) are separated by vertical
lines. It can be seen that the cluster assignment correlates loosely
with the final (radiation sensitive) cell type.}}

\newcommand{\prosAcap}{{
Fiedler vector values for spectral clustering of Singh prostate
data for three pathways: (a) complement and coagulation cascade,
(b) metabolism of xenobiotics by cytochrome P450, and (c) tryptophan metabolism. }
{Shown are each sample's Fiedler
vector value along with the resulting clustering.  A Gaussian mixture
fit to the density (left panel) of the Fiedler vector indicates
two clusters; the resulting cluster assignment for each sample is
indicated by color. True phenotype categories are
given as shapes: open circles denote non-tumor specimens; triangles, tumor.}}

\newcommand{\prosBcap}{{ Fiedler vector values for second PDM layer (scrubbed) clustering of Singh prostate data for three pathways: (a) bile acid synthesis, (b) glycerolipid metabolism, and (c) drug metabolism by cytochrome P450. }
{Shown are each sample's Fiedler
vector value along with the resulting clustering.  A Gaussian mixture
fit to the density (left panel) of the Fiedler vector indicates
two clusters; the resulting cluster assignment for each sample is
indicated by color. True phenotype categories are
given as shapes: open circles denote non-tumor specimens; triangles, tumor.}}

\newcommand{\prosbothcap}{{ 
Fiedler vector values for first PDM layer (a) and second (scrubbed) PDM layer (b) for the combined prostate data. } 
{Shown are each sample's Fiedler
vector value along with the resulting clustering.  A Gaussian mixture
fit to the density (left panel) of the Fiedler vector indicates
two clusters; the resulting cluster assignment for each sample is
indicated by color. True phenotype categories are
given as shapes: lower case `s' and `t' refer to stromal and normal samples
from the Singh~\cite{SING02} data, upper case `N', `S', `T', and `M' refer
to normal, stromal, tumor, and metastatic samples from the Yu data~\cite{YU04}.
}}

\newcommand{\twocirccap}{{two\_circles examples.} {In (a) and (b), colors denote cluster assignments from $k$-means ($k=2$) and spectral clustering, respectively. In (a), $k$-means using $k=2$ produces a linear cut through the data; in the (b), spectral clustering automatically chooses two clusters and assigns clusters with nonconvex boundaries. The embedded data used in (b) is shown in (c); in this representation, the clusters are linearly separable, and a rug plot shows the bimodal density of the Fiedler vector that yielded the correct number of clusters.}}

\newcommand{\spellcyccap}{{Yeast cell cycle data.}
{Expression levels for three oscillatory genes are shown.  The method of cell cycle synchronization is shown as shapes: crosses denote elutriation-synchronized samples, while triangles denote CDC-28 synchronized samples.  Cluster assignment for each sample is shown by color; above the diagonal, points are colored by $k$-means clustering, with poor correspondence between cluster (color) and synchronization protocol (shapes); below the diagonal, samples are colored by spectral clustering assignment, showing clusters that correspond to the synchronization protocol.}}

\newcommand{\twocircfiedcap}{{Laplacian matrix eigenvalues (top) and
Fiedler vector values (bottom) for spectral clustering of two\_circles
data.} {In the top plot, the resampling-based threshold for eigenvalue
significance is shown in cyan, with smaller eigenvalues plotted in
red.  In the bottom plot, we show each sample's Fiedler vector value
along with the resulting clustering.  A Gaussian mixture fit to the
density (bottom left) of the Fiedler vector indicates two clusters;
the resulting cluster assignment for each sample is indicated by
color. The true class labels (inner, outer ring) are given as
shapes, and it can be seen that the cluster assignment corresponds to
the class labels without error.}}

\ifthenelse{\boolean{withfigs}}
{
}{
        \clearpage 
        \section*{Figure Captions}
}

\flexfig{width=3in}{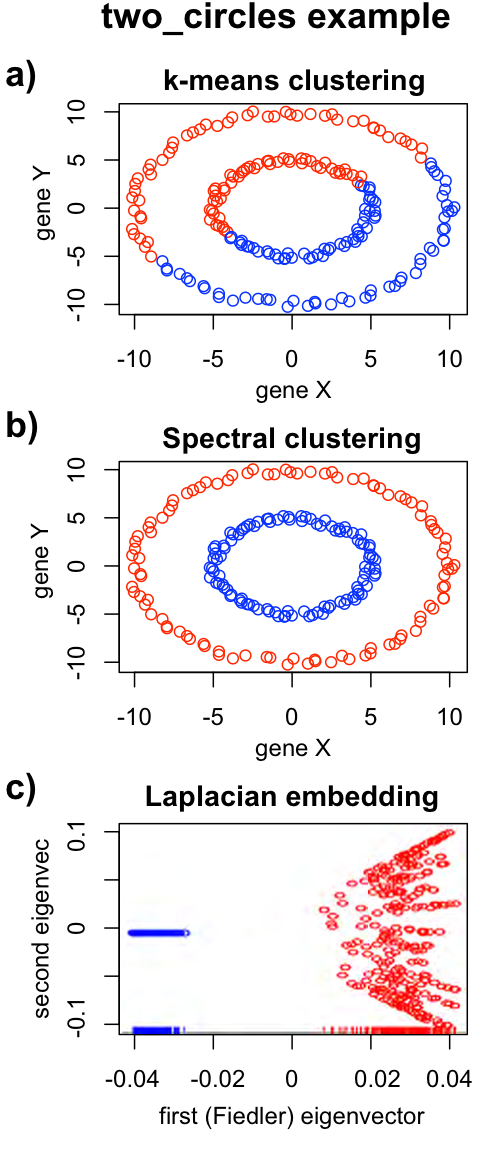}{twocircs}{\twocirccap}
\flexfig{width=6in}{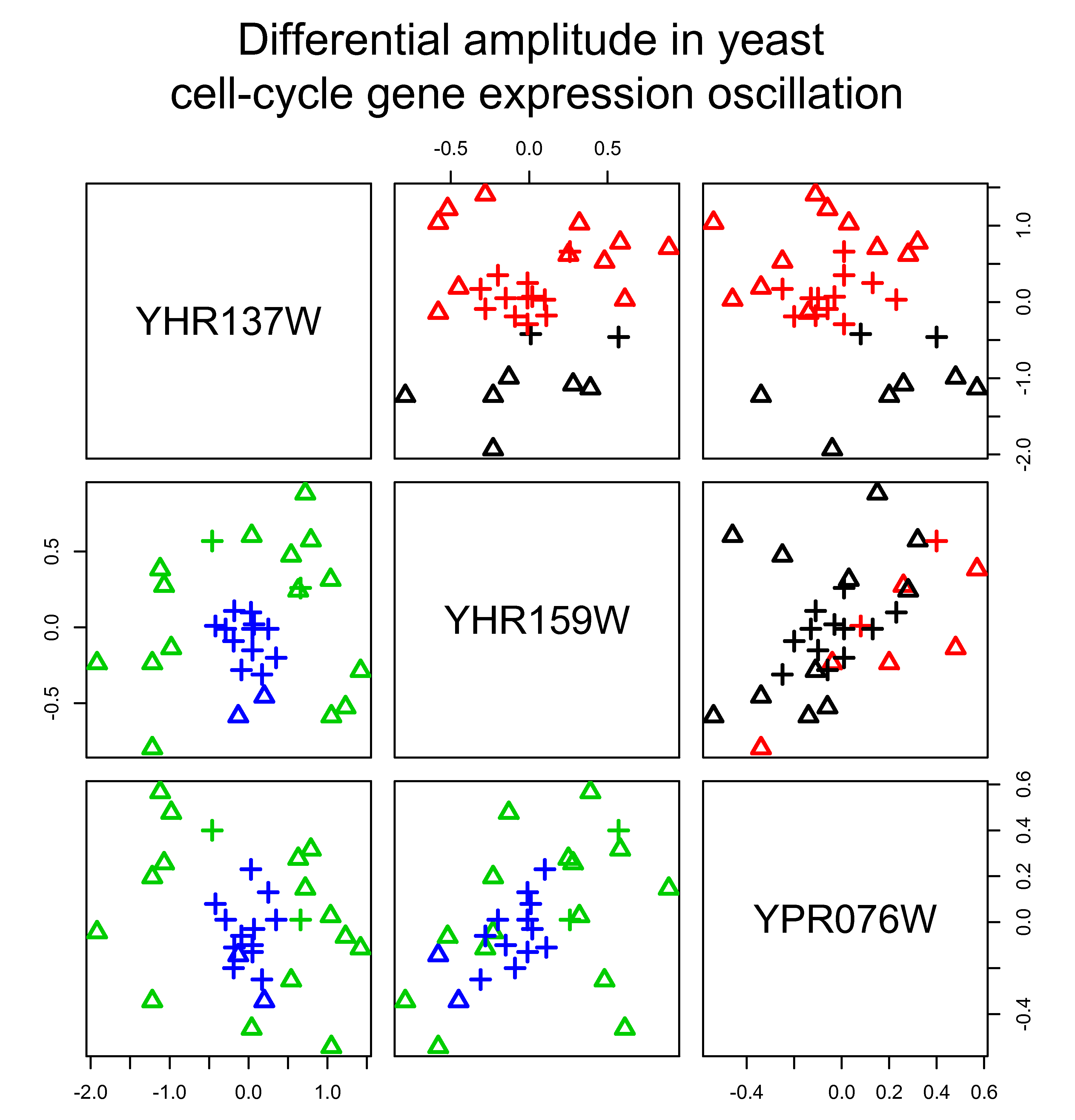}{spellcyc}{\spellcyccap}
\flexfig{width=3.5in}{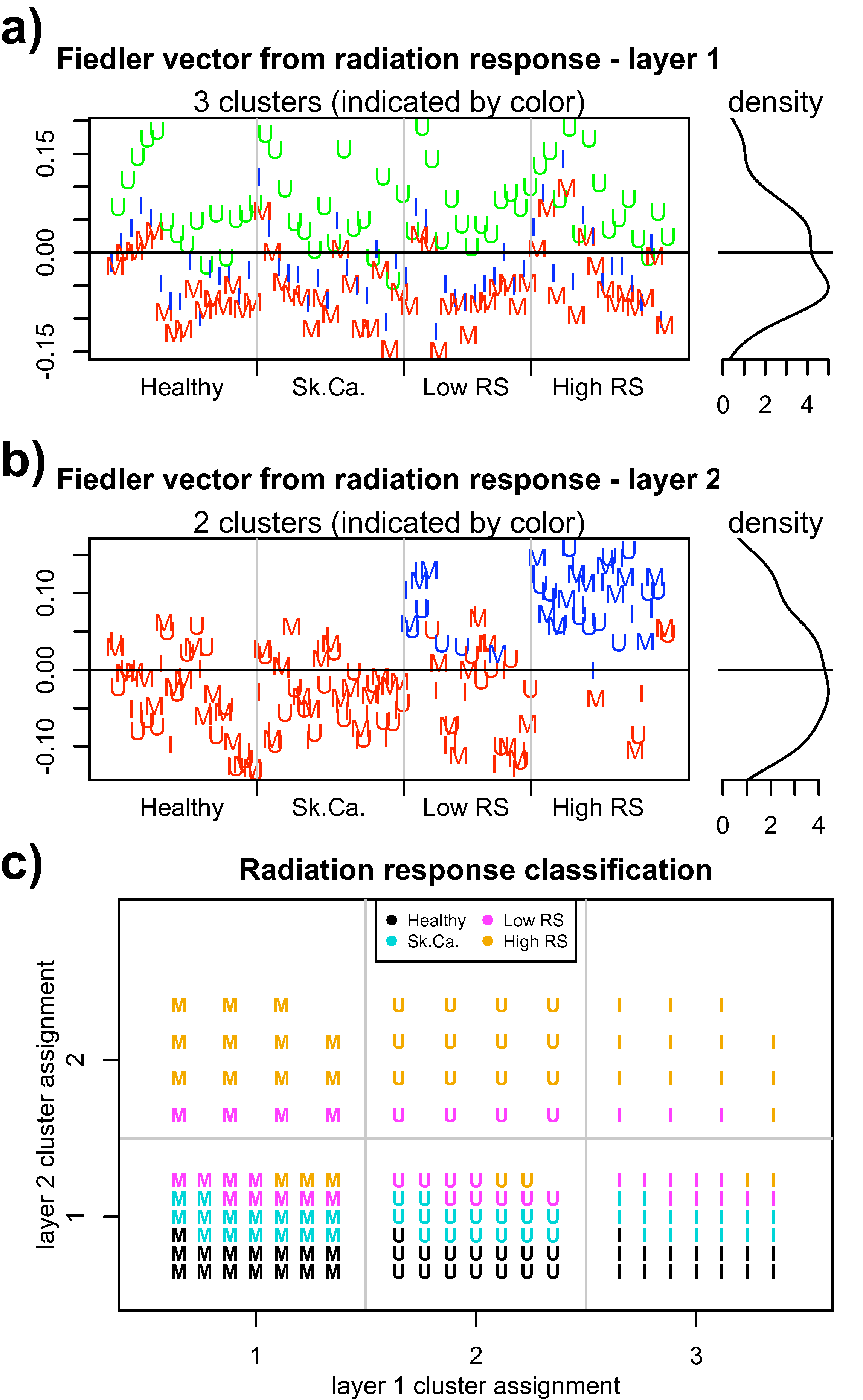}{pdmfig}{\pdmradcap}
\flexfig{width=6in}{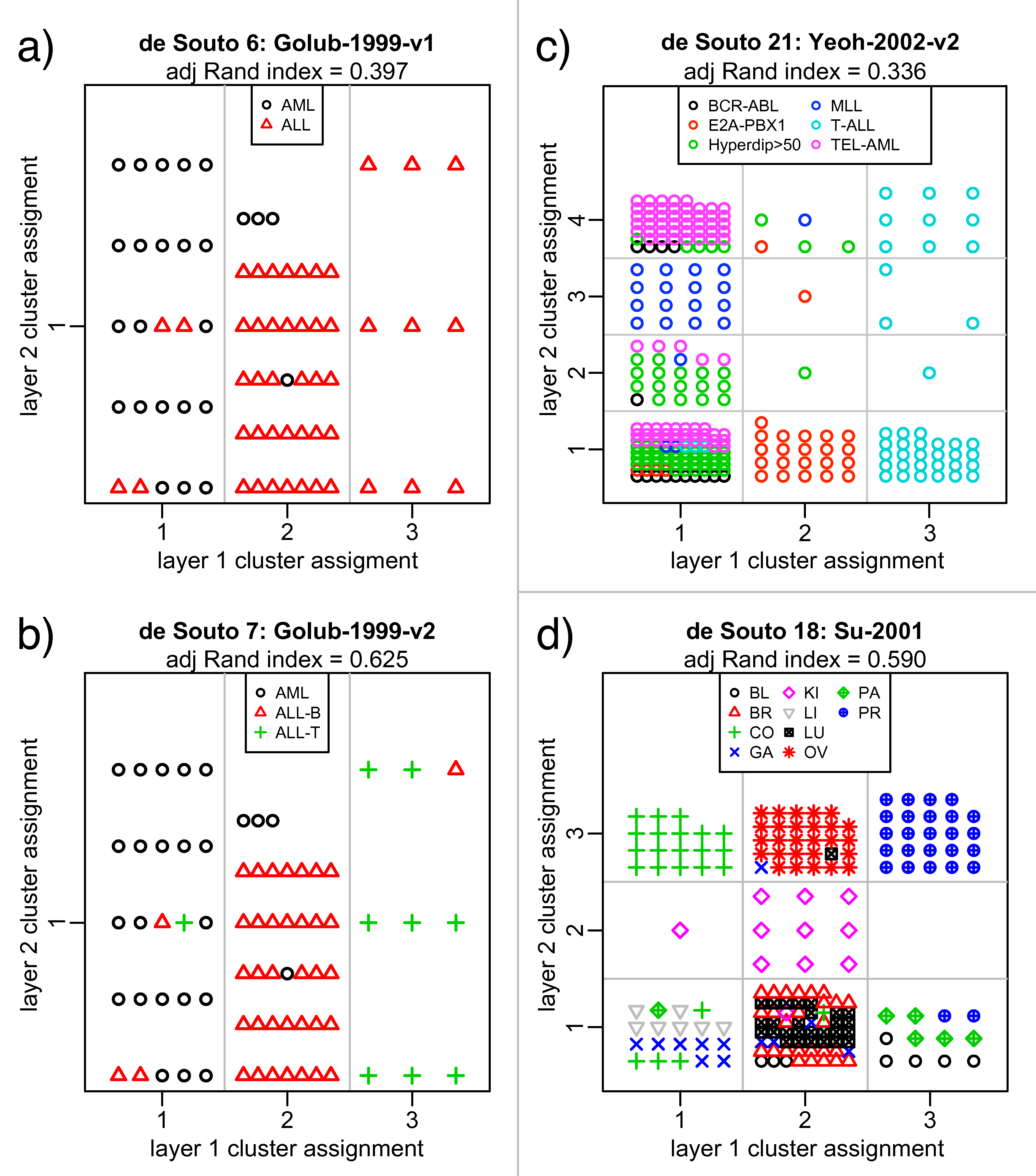}{disamb}{\disambcap}
\flexfig{width=6in}{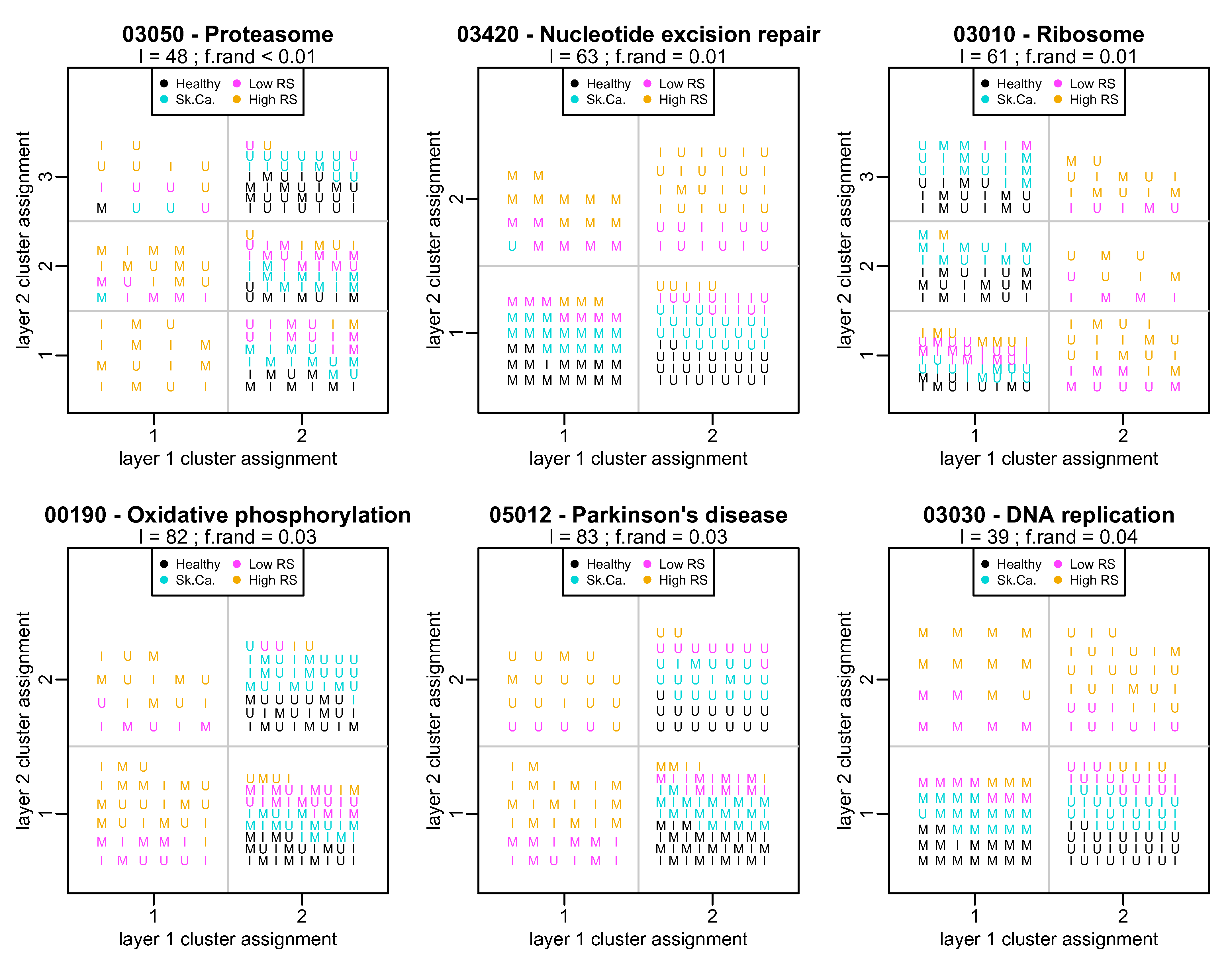}{radpath}{\radpathcap}
\flexfig{width=6in}{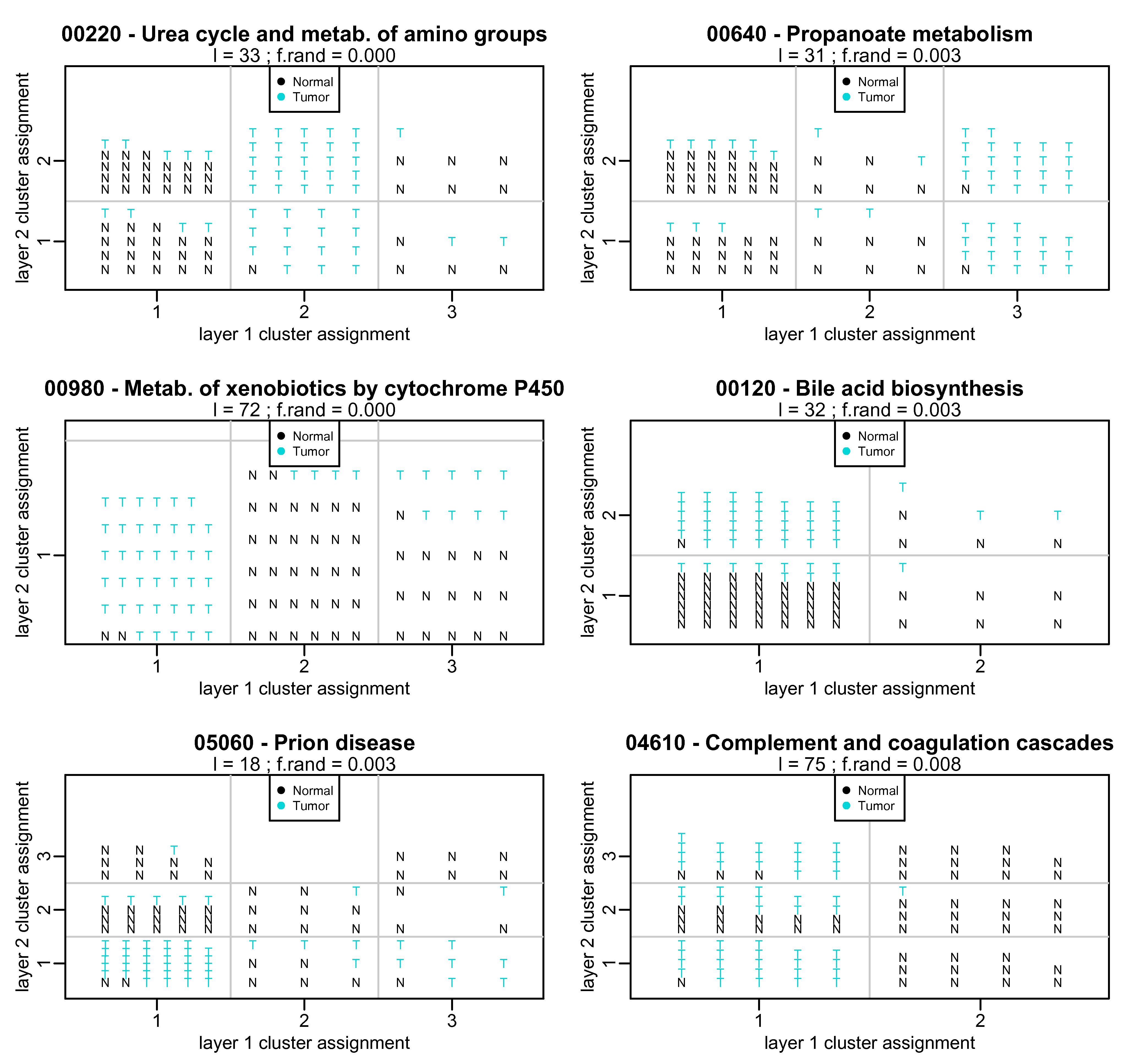}{prospath}{\prospathcap}


%
\renewcommand{\thefigure}{S-\arabic{figure}}
\renewcommand{\thetable}{S-\arabic{table}}
\setcounter{figure}{0}
\setcounter{table}{0}

\setboolean{withfigs}{false}

\clearpage
\section*{Supplemental figures}

\flexfig{width=6in}{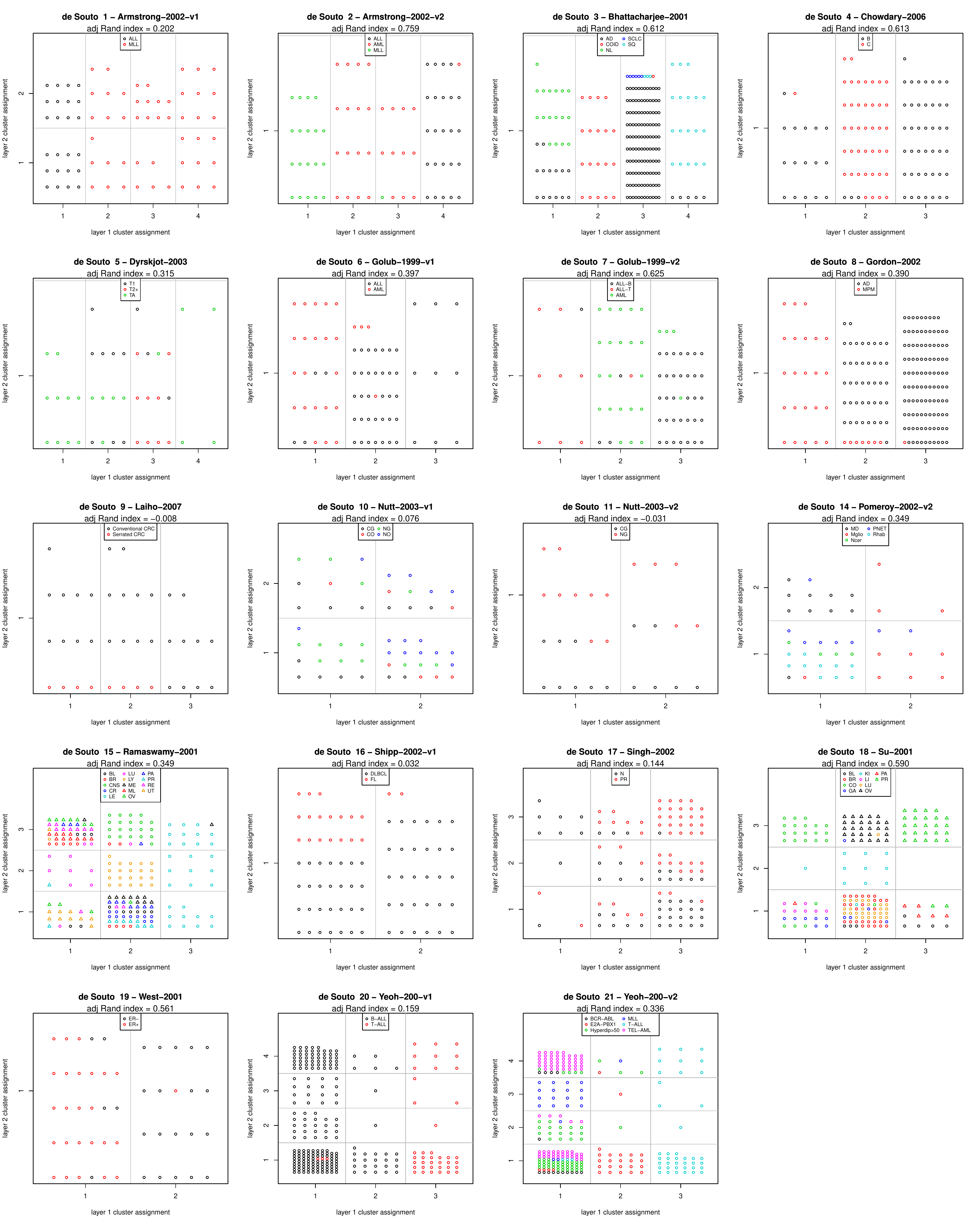}{deso}{PDM classifications of deSouto benchmark set samples using a correlation-based distance metric (as described in methods).}
\flexfig{width=6in}{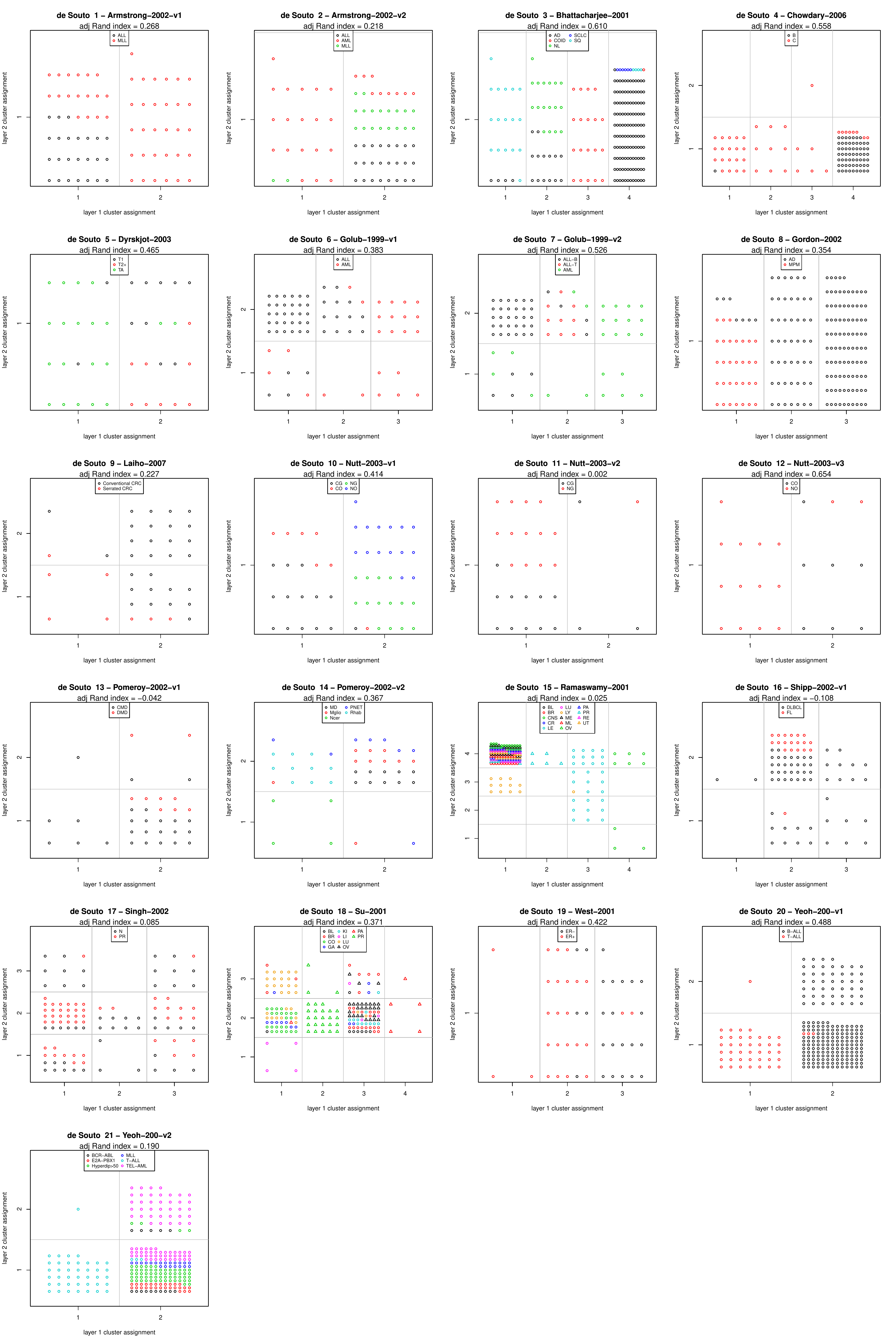}{desoeuc}{PDM classifications of deSouto benchmark set samples using a Euclidean distance metric.}
\flexfig{width=6in}{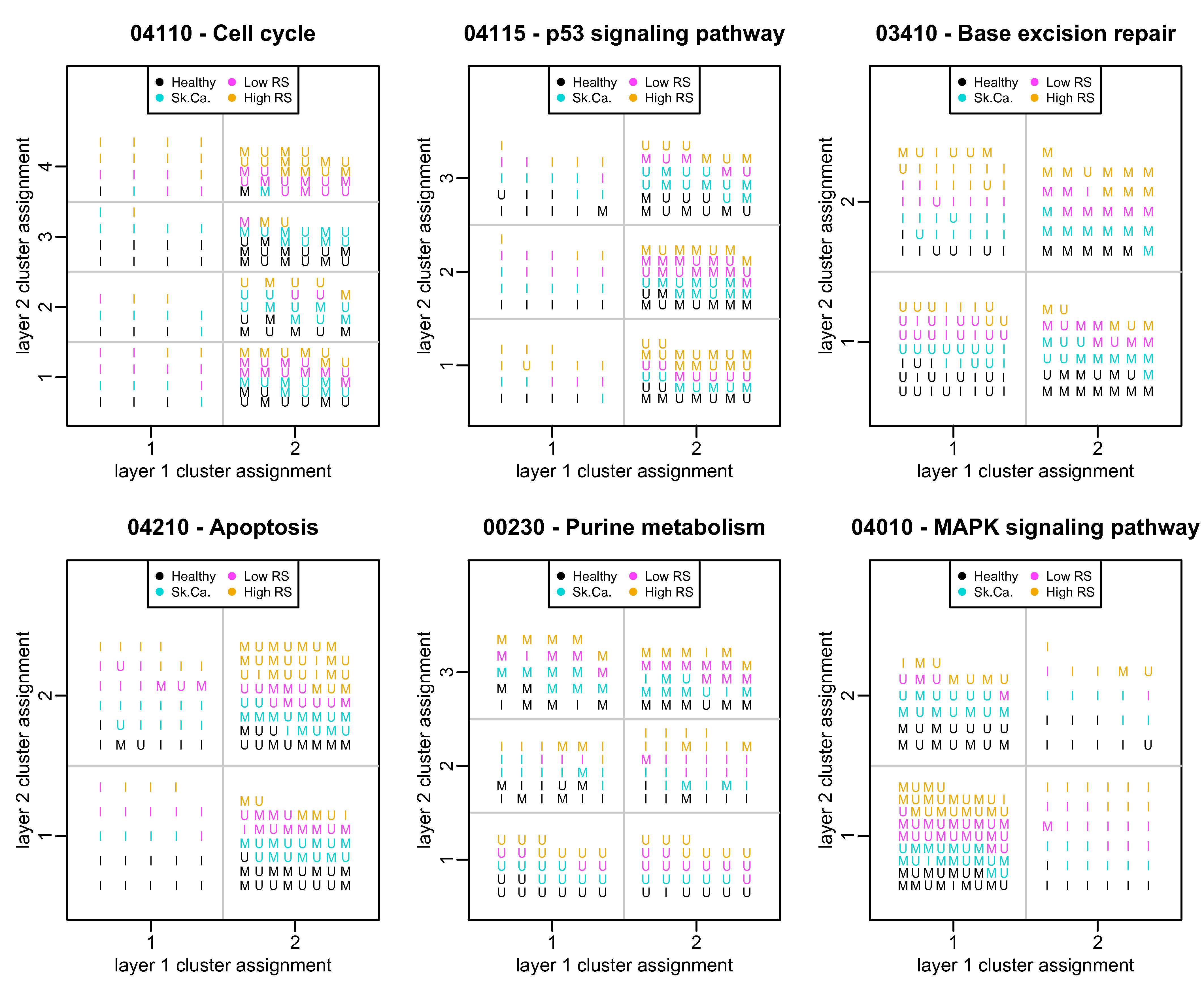}{skp2}{Pathway-PDM classifications of radiation response data for pathways that discriminate cells by radiation exposure but not by phenotype, suggesting that these mechanisms are intact across sample types.  Exposure is indicated by 
shape (``M'', mock; ``U'', UV; ``I'', IR), with phenotypes (healthy, skin cancer, low RS, high RS) indicated by color. The discriminatory pathways relate to DNA metabolism and cell death, as would be expected from radiation exposure.}
\flexfig{width=3in}{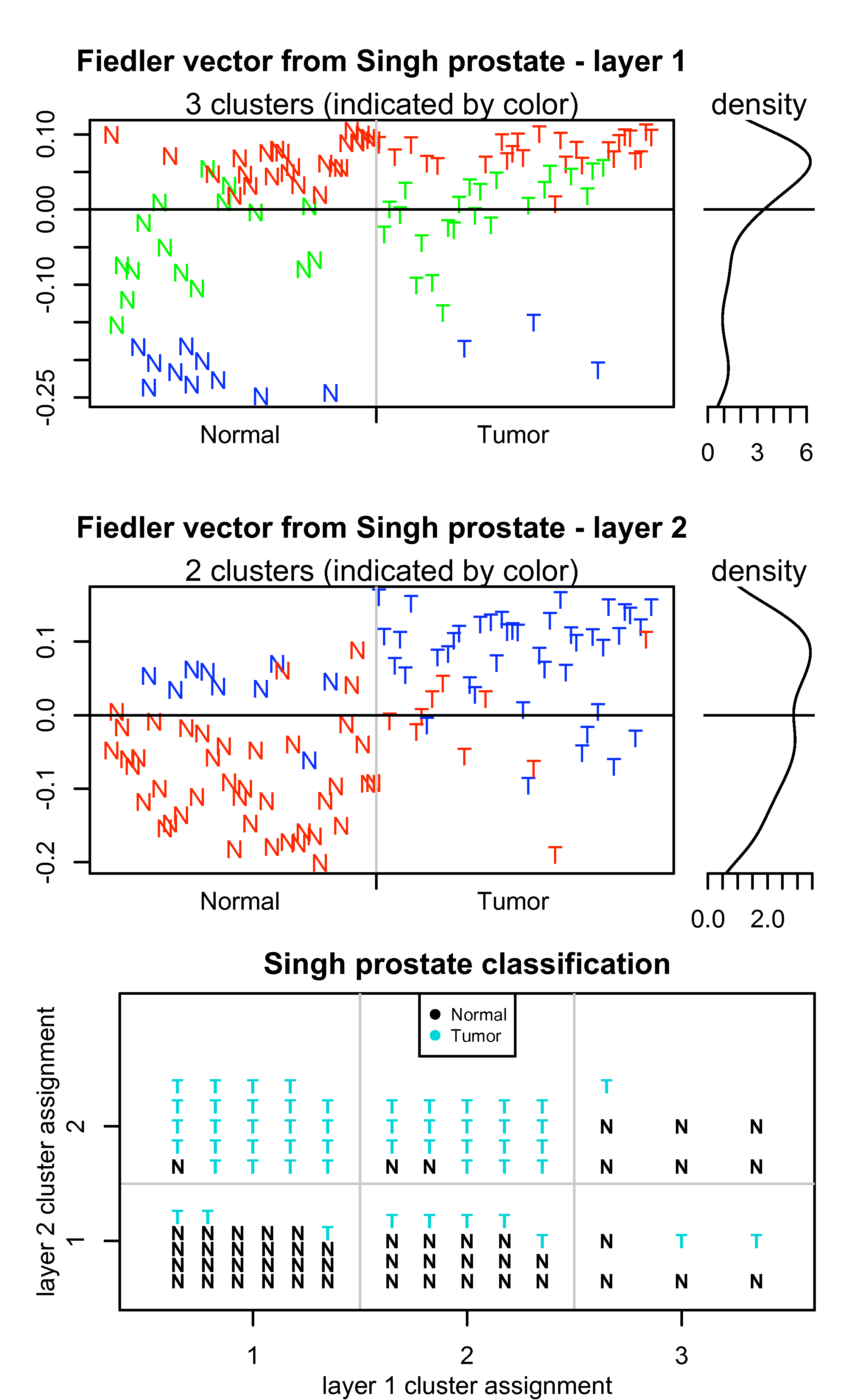}{prospdm}{PDM results in first and second layers of the Singh prostate tumor data using all genes.  The top two panels show the Fiedler vector values and clustering results, along with the Fiedler vector density, in the first and second layer; the bottom panel shows the combined classification results.  The second layer, but not the first, discriminates the tumor samples.}

\end{document}